\documentclass[11pt,onecolumn,dvips,draftcls]{IEEEtran}

\usepackage{psfig,amsfonts,amsmath,color,amssymb,amsxtra}
\usepackage[breaklinks=true, colorlinks=true, linkcolor=black,
urlcolor=dblue, citecolor=black, pdfpagemode=None, pdfstartview=FitH]{hyperref}
\DeclareOldFontCommand{\rm}{\normalfont\rmfamily}{\mathrm}
\DeclareOldFontCommand{\sf}{\normalfont\sffamily}{\mathsf}
\DeclareOldFontCommand{\tt}{\normalfont\ttfamily}{\mathtt}
\DeclareOldFontCommand{\bf}{\normalfont\bfseries}{\mathbf}
\DeclareOldFontCommand{\it}{\normalfont\itshape}{\mathit}
\DeclareOldFontCommand{\sl}{\normalfont\slshape}{\@nomath\sl}
\DeclareOldFontCommand{\sc}{\normalfont\scshape}{\@nomath\sc}
\definecolor{gray}{cmyk}{.2,0.2,.3,.1}
\definecolor{dred}{cmyk}{0,0.9,0.4,0.3}
\definecolor{dblue}{rgb}{0,0,0.5}
\definecolor{dgreen}{rgb}{0,0.3,0}
\definecolor{dgray}{rgb}{0.3,0.3,0}
\newtheorem{theorem}{Theorem}
\newtheorem{lemma}{Lemma}
\newtheorem{corollary}{Corollary}

\newcommand{\tend}{\hfill$\blacksquare$}

\newcommand{\flow}{\varphi}
\title{Network Information Flow \\ with Correlated Sources
  \thanks{J.\ Barros was with the
  Institute for Communications Engineering, Munich University of Technology,
  Munich, Germany.  He is now with the Department of Computer Science,
  University of Porto, Portugal.  URL:
  \href{http://www.dcc.fc.up.pt/\~barros/}
  {{\tt http://www.dcc.fc.up.pt/$\sim$barros/}}.
  S.\ D.\ Servetto is with
  the School of Electrical and Computer Engineering, Cornell University,
  Ithaca, NY.  URL:
  \href{http://cn.ece.cornell.edu/}{{\tt http://cn.}}
  \href{http://cn.ece.cornell.edu/}{{\tt ece.cornell.edu/}}.
  Work supported by a scholarship from
  the Fulbright commission, and by the National Science Foundation, under
  awards CCR-0238271 (CAREER), CCR-0330059, and ANR-0325556.  Previous
  conference publications:~\cite{BarrosS:02b, BarrosS:03a, BarrosS:03d,
  BarrosS:05}.}}
\author{Jo\~{a}o Barros \hspace{2cm} Sergio D.\ Servetto}
\date{October 2, 2005.}

\begin{document}
\maketitle

\begin{picture}(0,0)
\put(0,220){\tt\small To appear in the IEEE Transactions on Information
  Theory.}
\end{picture}

\vspace{-13mm}
\begin{abstract}
\noindent\it
Consider the following network communication setup, originating
in a sensor networking application we refer to as the ``sensor
reachback'' problem.  We have a directed graph $G=(V,E)$, where
$V = \{v_0v_1...v_n\}$ and $E\subseteq V\times V$.  If $(v_i,v_j)\in E$,
then node $i$ can send messages to node $j$ over a discrete memoryless
channel $(\mathcal{X}_{ij},p_{ij}(y|x),\mathcal{Y}_{ij})$, of capacity
$C_{ij}$.  The channels are independent.  Each node $v_i$ gets to
observe a source of information $U_i$ ($i=0...M$), with joint
distribution $p(U_0U_1...U_M)$.  Our goal is to solve an incast
problem in $G$: nodes exchange messages with their neighbors, and after
a finite number of communication rounds, one of the $M+1$ nodes ($v_0$
by convention) must have received enough information to reproduce
the entire field of observations $(U_0U_1...U_M)$, with arbitrarily
small probability of error.  In this paper, we prove that such perfect
reconstruction is possible if and only if
\[
  H(U_S|U_{S^c}) \;\;<\;\; \sum_{i\in S,j\in S^c} C_{ij},
\]
for all $S \subseteq \{0...M\}$, $S\neq\emptyset$, $0\in S^c$.  Our
main finding is that in this setup a general source/channel separation
theorem holds, and that Shannon information behaves as a classical
network flow, identical in nature to the flow of water in pipes.  At
first glance, it might seem surprising that separation holds in a
fairly general network situation like the one we study.  A closer look,
however, reveals that the reason for this is that our model allows
only for independent point-to-point channels between pairs of nodes,
and not multiple-access and/or broadcast channels, for which separation
is well known not to hold~\cite[pp.\ 448-49]{CoverT:91}.  This
``information as flow'' view provides an algorithmic interpretation
for our results, among which perhaps the most important one is the
optimality of implementing codes using a {\em layered} protocol stack.
\end{abstract}

\vspace{1cm}
\pagebreak

\section{Introduction}

\subsection{The Sensor Reachback Problem}

\begin{figure}[ht]
\centerline{\psfig{width=12cm,height=4cm,file=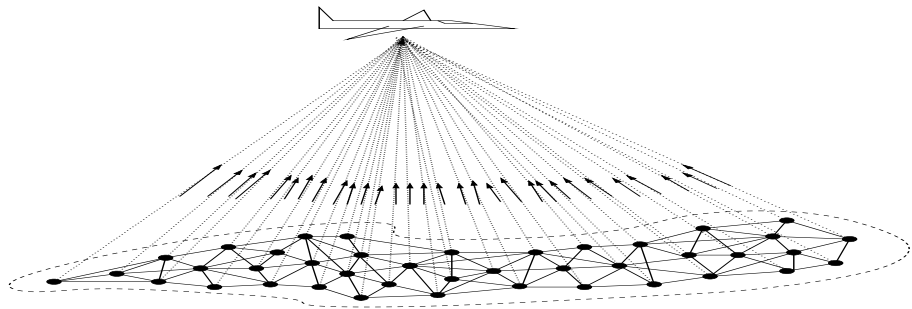}}
\caption{A large number of sensors is deployed over a target area.
  After collecting the data of interest, the sensors must {\it reach back}
  and transmit this information to a single receiver (e.g., an overflying
  plane) for further processing.}
\label{fig:reachback}
\end{figure}

Wireless sensor networks made up of small, cheap, and mostly unreliable
devices equipped with limited sensing, processing and transmission
capabilities, have recently sparked a fair amount of interest in
communications problems involving multiple correlated sources and
large-scale wireless networks~\cite{AkyildizSSC:02}.  It is envisioned
that an important class of applications for such networks involves a
dense deployment of a large number of sensors over a fixed area, in
which a physical process unfolds---the task of these sensors is then
to collect measurements, encode them, and relay them to some data
collection point where this data is to be analyzed, and possibly acted
upon.  This scenario is illustrated in Fig.~\ref{fig:reachback}.

There are several aspects that make this communications problem
interesting:
\begin{itemize}
\item {\it Correlated Observations:} If we have a large number of nodes
  sensing a physical process within a confined area, it is reasonable to
  assume that their measurements are correlated. This correlation may be
  exploited for efficient encoding/decoding.
\item {\it Cooperation among Nodes:} Before transmitting data to the
  remote receiver, the sensor nodes may establish a {\it conference}
  to exchange information over the wireless medium and increase their
  efficiency or flexibility through cooperation.
\item {\it Channel Interference:} If multiple sensor nodes use the wireless
  medium at the same time (either for conferencing or reachback), their
  signals will necessarily interfere with each other.  Consequently,
  reliable communication in a reachback network requires a set of rules
  that control (or exploit) the interference in the wireless medium.
\end{itemize}

In order to capture some of these key aspects, while still being able
to provide complete results, we make some modeling assumptions, discussed
next.

\subsubsection{Source Model}

We assume that the sources are memoryless, and thus consider only the
spatial correlation of the observed samples and not their temporal
dependence (since the latter dependencies could be dealt with by simple
extensions of our results to the case of ergodic sources).  Furthermore,
each sensor node $v_i$ observes only one component $U_i$ and must
transmit enough information to enable the sink node $v_0$ to reconstruct
the whole vector $U_1U_2\dots U_M$.  This assumption is the most natural
one to make for scenarios in which data is required at a remote location
for fusion and further processing, but the data capture process is
distributed, with sensors able to gather {\em local} measurements only,
and deeply embedded in the environment.

A conceptually different approach would be to assume that all sensor
nodes get to observe independently corrupted noisy versions of one and
the same source of information $U$, and it is this source (and not the
noisy measurements) that needs to be estimated at a remote location.
This approach seems better suited for applications involving non-homogeneous
sensors, where each one of the sensors gets to observe different
characteristics of the same source (e.g., multispectral imaging), and
therefore leads to a conceptually very different type of sensing
applications from those of interest in this work.  Such an approach
leads to the so called {\it CEO problem} studied by Berger, Zhang and
Viswanathan in~\cite{BergerZV:96}.

\subsubsection{Independent Channels}

Our motivation to consider a network of independent DMCs is twofold.

From a pure information-theoretic point of view independent channels
are interesting because, as shown in this paper, this assumption gives
rise to long Markov chains which play a central role in our ability to
prove the converse part of our coding theorem, and thus obtain conclusive
results in terms of capacity.  Moreover, a corollary of said coding
theorem does provide a conclusive answer for a special case of the
multiple access channel with correlated sources, a problem for which
no general converse is known.

From a more practical point of view, the assumption of independent
channels is valid for any network that controls interference by means
of a reservation-based medium-access control protocol (e.g., TDMA).
This option seems perfectly reasonable for sensor networking scenarios
in which sensors collect data over extended periods of time, and must
then transmit their accumulated measurements simultaneously.  In this
case, a key assumption in the design of standard random access techniques
for multiaccess communication breaks down---the fact that individual
nodes will transmit with low probability~\cite[Chapter~4]{BertsekasG:92}.
As a result, classical random access would result in too many collisions
and hence low throughput.  Alternatively, instead of {\em mitigating}
interference, a medium access control (MAC) protocol could attempt to
{\em exploit} it, in the form of using cooperation among nodes to generate
waveforms that add up constructively at the receiver (cf.~\cite{HuS:03c,
HuS:03b, HuS:05}).  Providing an information-theoretic analysis of such
cooperation mechanisms would be very desirable, but since it entails
dealing with correlated sources and a general multiple access channel,
dealing with correlated sources and an array of independent channels
constitutes a reasonable first step towards that goal, and is also
interesting in its own right, since it provides the ultimate performance
limits for an important class of sensor networking problems.

\subsubsection{Perfect Reconstruction at the Receiver}

In our formulation of the sensor reachback problem, the far receiver
is interested in reconstructing the entire field of sensor measurements
with arbitrarily small probability of error.  This formulation leads
us to a natural {\em capacity} problem, in the classical sense of
Shannon.  Alternatively, one could relax the condition of perfect
reconstruction, and tolerate some distortion in the reconstruction
of the field of measurements at the far receiver, thus leading to
the so called {\em Multiterminal Source Coding} problem studied by
Berger~\cite{Berger:78}.  This condition could be further relaxed,
to require a faithful reproduction of the {\em image} of some function
$f$ of the sources, leading to a problem studied extensively by
Csiszar, K\"orner and Marton~\cite{CsiszarK:80, KoernerM:79}.

\subsection{An Information Theoretic View of Architectural Issues}

For large-scale, complex systems of the type of interest in this work,
the complexity of basic questions of design and performance analysis
appears daunting:
\begin{itemize}
\item How should nodes cooperate to relay messages to the data collector
  node $v_0$?  Should they decode received messages, re-encode them, and
  forward to other nodes?  Should they map channel outputs to channel
  inputs without attempting to decode?  Should they do something else?
\item How should redundancy among the sources be exploited?  Should we
  compress the information as much as possible?  Should we leave some
  of that redundancy to combat noise in the channels?  Is there a
  source/channel separation theorem in these networks?
\item How do we measure performance of these networks, what are appropriate
  cost metrics?  How do we design networks that are efficient under an
  appropriate cost metric?
\end{itemize}
In~\cite{KawadiaK:04}, a number of examples are identified in which
the existence of a simple architecture has played an enabling role in
the proliferation of technology: the von Neuman computer architecture,
separation of source and channel coding in communications, separation
of plant and controller in control systems, and the OSI layered
architecture model.  So what all these questions boil down to is
an issue similar to those considered in~\cite{KawadiaK:04}: what are
appropriate abstractions of the network, similar to the IP protocol
stack for the Internet, based on which we can break the design task
into independent reusable components, optimize the design of these
components, and obtain an {\em efficient} system as a result?  In
this work, we show how information theory is indeed capable of
providing very meaningful answers to this problem.

\begin{figure}[!ht]
\centerline{\psfig{file=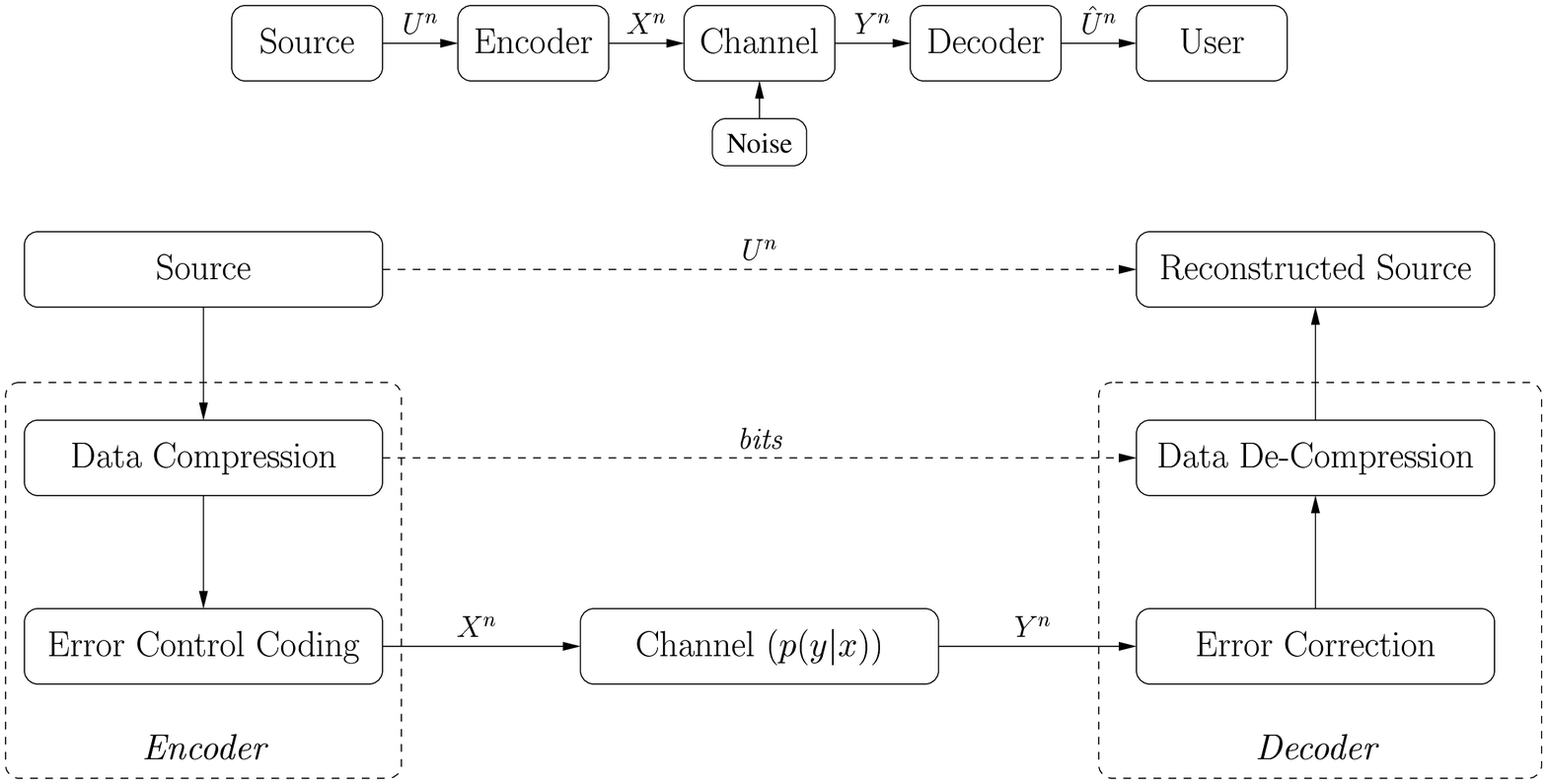,height=5cm,width=14cm}}
\caption{\small Shannon's model for a point-to-point system.  Top
  figure: abstract view, consisting of a source, an
  encoder from source symbols to channel symbols, a conditional
  probability distribution to model the random dependence of outputs
  on inputs, and a decoder to map from received messages back to
  source symbols; bottom figure: a capacity-achieving architecture
  for this system, in which error control codes are used to create
  an illusion of a noiseless bit pipe.}
\label{fig:shannon-pt2pt}
\end{figure}

Information theory, in one of its applications, deals with the analysis
of performance of communication systems.  So, to some it may seem the
natural theory to turn to for guidance in the task of searching for a
suitable network architecture.  However, to others it may seem unnatural
to do so: it is well known that information theory and communication
networks have not had fruitful interactions in the past, as explained
by Ephremides and Hajek~\cite{EphremidesH:98}.  Thus, in the presence
of these mixed indicators, we take the stand that indeed information
theory has a great deal to offer in the task at hand.  And to justify
our position, consider Shannon's model for a communications system, as
illustrated in Fig.~\ref{fig:shannon-pt2pt}.
 
For this setup, Shannon established that reliable communication of
a source over a noisy channel is possible if and only if the entropy
rate of the source is less than the capacity of the
channel~\cite[Ch.\ 8.13]{CoverT:91}.  This result, known as the
source/channel separation theorem, has a double significance.  On
one hand, it provides an exact single-letter characterization of
conditions under which reliable communication is possible.  On the
other hand, and of particular interest to the task at hand for
us, it is a statement about the {\em architecture} of an optimal
communication system: the encoder/decoder design task can be split
into the design and optimization of two independent components.
So it is inspired by Shannon's teachings for point-to-point
systems that we ask in this work, and answer in the affirmative,
the question of whether it is possible or not to derive similar
useful architectural guidelines for the class of networks under
consideration.

\subsection{Related Work}
\label{sec:related-work}

The problem of communicating distributed correlated sources over a
network of point-to-point links is closely related to several classical
problems in network information theory.  To set the stage for the main
contributions of this paper, we now review related previous work.

\subsubsection{Distributed Correlated Sources and Multiple Access}

The concept of separate encoding of correlated sources was studied by
Slepian and Wolf in their seminal paper~\cite{SlepianW:73b},
where they proved that two correlated sources $(U_1U_2)$ drawn i.i.d.\
$\sim p(u_1u_2)$ can be compressed at rates $(R_1,R_2)$ if and only if
\begin{eqnarray*}
R_1 & \geq & H(U_1|U_2) \\
R_2 & \geq & H(U_2|U_1) \\
R_1+R_2 & \geq & H(U_1U_2).
\end{eqnarray*}

Assume now that $(U_1U_2)$ are to be transmitted with arbitrarily
small probability of error to a joint receiver over a multiple access
channel with transition probability $p(y|x_1x_2)$.  Knowing that the
capacity of the multiple access channel with independent
sources is given by the convex hull of the set of points $(R_1,R_2)$
satisfying~\cite[Ch.\ 14.3]{CoverT:91}
\begin{eqnarray*}
R_1&<&I(X_1;Y|X_2) \\
R_2&<&I(X_2;Y|X_1) \\
R_1+R_2&<&I(X_1X_2;Y),
\end{eqnarray*}
it is not difficult to prove that Slepian-Wolf source coding of
$(U_1U_2)$ followed by separate channel coding yields the following
{\em sufficient} conditions for reliable communication
\begin{eqnarray*}
H(U_1|U_2) & < & I(X_1;Y|X_2) \\
H(U_2|U_1) & < & I(X_2;Y|X_1) \\
H(U_1U_2) & < & I(X_1X_2;Y).
\end{eqnarray*}
These conditions, which basically state that the Slepian-Wolf region
and the capacity region of the multiple access channel have a non-empty
intersection, are sufficient but not necessary for reliable communication,
as shown by Cover, El Gamal, and Salehi with a simple counterexample
in~\cite{CoverGS:80}.  In that same paper, the authors introduce a class
of {\it correlated} joint source/channel codes, which enables them to
increase the region of achievable rates to
\begin{eqnarray}
H(U_1|U_2)&<&I(X_1;Y|X_2U_2) \label{eq:cegs1}\\
H(U_2|U_1)&<&I(X_2;Y|X_1U_1) \label{eq:cegs2}\\
H(U_1U_2)&<&I(X_1X_2;Y), \label{eq:cegs3}
\end{eqnarray}
for some
$p(u_1u_2x_1x_2y)=p(u_1u_2)\cdot p(x_1|u_1)\cdot p(x_2|u_2)\cdot p(y|x_1x_2)$.
Also in~\cite{CoverGS:80}, the authors generalize this set of sufficient
conditions to sources $(U_1U_2)$ with a common part $W=f(U_1)=g(U_2)$,
but they were not able to prove a converse, i.e., they were not able
to show that their region is indeed the capacity region of the multiple
access channel with correlated sources.  Later, it was shown with a
carefully constructed example by Dueck in~\cite{Dueck:81} that indeed
the region defined by eqns.~(\ref{eq:cegs1})-(\ref{eq:cegs3}) is not
tight.  Related problems were considered by Slepian and
Wolf~\cite{SlepianW:73}, and Ahlswede and Han~\cite{AhlswedeH:83}.  
To this date however, the general problem still
remains open.

Assuming independent sources, Willems investigated a cooperative
scenario, in which encoders exchange messages over {\em conference}
links of limited capacity prior to transmission over the multiple
access channel~\cite{Willems:83}.  In this case, the capacity region
is given by
\begin{eqnarray*}
R_1&<&I(X_1;Y|X_2Z)+C_{12} \\
R_2&<&I(X_2;Y|X_1Z)+C_{21} \\
R_1+R_2&<&\min\{\;I(X_1X_2;Y|Z)+C_{12}+C_{21},\;I(X_1X_2;Y)\;\},
\end{eqnarray*}
for some auxiliary random variable $Z$ such that
$|\mathcal{Z}|\leq\min(|\mathcal{X}_1|\cdot|\mathcal{X}_2|+2,|\mathcal{Y}|+3)$,
and for a joint distribution $p(zx_1x_2y_1y_2)
= p(z)\cdot p(x_1|z)\cdot p(x_2|z)\cdot p(y|x_1x_2)$.

\subsubsection{Correlated Sources and Networks of DMCs}

Very recently, an early paper was brought to our attention, in which 
Han considers the transmission of correlated sources to a common sink
over a network of independent channels~\cite{Han:80}.  Although the
problem setup is less general than ours, in that (a) each source block
and each transmitted codeword partipate only once in the encoding
process, and (b) the intermediate nodes are assumed to decode the
data before passing it on, Theorem 3.1 of~\cite{Han:80} is very similar
to our Theorem~\ref{thm:main}.

Our work, done independently of Han's, differs from it and complements
it in the following ways:
\begin{itemize}
\item Our setup is more general.  We allow for arbitrary forms of joint
  source-channel coding to take place inside the network while data flows
  towards the decoder, and then {\em prove} that a one-step encoding
  process, pure routing, and separate source/channel coding are sufficient.
  Han assumes decode-and-forward in his problem statement, as well as
  a one-step encoding process.
\item The proof techniques are different.  Han takes a purely combinatorial
  approach to the problem: he thoroughly exploits the polymatroidal
  structure of the capacity function for the network of channels, and the
  co-polymatroidal structure for the Slepian-Wolf region.  We establish our
  achievability result by explicitly constructing a routing algorithm for
  the Slepian-Wolf indices, and our converse by standard methods based on
  Fano's inequality.
\end{itemize}
Furthermore our work, being motivated by a concrete sensor networking
application, establishes connections and relevance to practical engineering
problems (see Section~\ref{sec:protocol-stack}) that are not a concern
in~\cite{Han:80}.

\subsubsection{Network Coding}

Another closely related problem is the well known {\em network coding}
problem, introduced by Ahlswede, Cai, Li and Yeung~\cite{AhlswedeCLY:00}.
In that work, the authors establish the need for applying coding
operations at intermediate nodes to achieve the max-flow/min-cut bound
of a general multicast network.  A converse proof for this problem
was provided by Borade~\cite{Borade:02}.  Linear codes were proposed
by Li, Yeung and Cai in~\cite{LiYC:03}, and Koetter and M\'edard
in~\cite{KoetterM:03}.

Effros, M\'edard et al.\ have developed a comprehensive study on separate
and joint design of linear source, channel and network codes for networks
with correlated sources under the assumption that all operations are
defined over a common finite field~\cite{EffrosMHRKK:03}.  For this
particular case, optimality of separate linear source and channel coding
was observed in the one-receiver instance, but the result
of~\cite{EffrosMHRKK:03} does not prove that it holds for general networks
and channels with arbitrary input and output alphabets.  Error exponents
for multicasting of correlated sources over a network of noiseless channels
were given by Ho, M\'edard et al.\ in~\cite{HoMEK:04}, and networks with
undirected links were considered by Li and Li in~\cite{LiL:04}.

Another problem in which network flow techniques have been found useful
is that of finding the maximum stable throughput in certain networks.  In
this problem, posed by Gupta and Kumar in~\cite{GuptaK:00}, it is sought
to determine the maximum rate at which nodes can inject bits into a
network, while keeping the system stable.  This problem was reformulated
by Peraki and Servetto as a multicommodity flow problem, for which tight
bounds were obtained using elementary counting
techniques~\cite{PerakiS:03, PerakiS:04}.

\subsection{Main Contributions and Organization of the Paper}

Our main original contributions can be summarized as follows:
\begin{itemize}
\item A general coding theorem yielding necessary and sufficient
  conditions for reliable communication of $M+1$ correlated sources
  to a common sink over a network of independent DMCs.
\item An achievability proof which combines classical coding arguments
  with network flow methods and a converse proof that establishes
  the optimality of separate source and channel coding. 
\item A detailed discussion on the engineering implications of our
  main result, and the concepts of information-theoretically optimal
  network architectures and protocol stacks.
\end{itemize}

The rest of the paper is organized as follows.  In
Section~\ref{sec:coding-theorems} we give formal definitions, to then
state and prove our main theorem.  We also look at three special cases:
a network with three nodes, the non-cooperative case, and an array of
orthogonal Gaussian channels.  In Section~\ref{sec:protocol-stack} we
address the practical implications of our main result, by describing
an information-theoretically optimal protocol stack, elaborating on the
tractability of related network architecture and network optimization
problems, and discussing the suboptimality of correlated codes for
orthogonal channels.  The paper concludes with
Section~\ref{sec:conclusions}.

\section{A Coding Theorem for Network Information Flow with Correlated
  Sources}
\label{sec:coding-theorems}

\subsection{Formal Definitions and Statement of the Main Theorem}

A {\em network} is modeled as the complete graph on $M+1$ nodes.
For each $(v_i,v_j)\in E$ ($0\leq i,j\leq M$), there is a discrete
memoryless channel $(\mathcal{X}_{ij},p_{ij}(y|x),\mathcal{Y}_{ij})$,
with capacity $C_{ij} = \max_{p_{ij}(x)} I(X_{ij};Y_{ij})$.\footnote{Note
that $C_{ij}$ could potentially be zero, thus assuming a complete graph
does not mean necessarily that any node can send messages to any other
node in one hop.}
At each node $v_i\in V$, a random variable $U_i$ is observed
($i=0...M$), drawn i.i.d.\ from a known joint distribution
$p(U_0U_1...U_M)$.  Node $v_0$
is the {\em decoder} -- the goal in this problem is to
find conditions under which $U_1...U_M$ can be reproduced reliably at
$v_0$.  We now make this statement more precise, by describing how the
nodes communicate and by giving the formal definitions of code,
probability of error and reliable communication.

Time is discrete.  Every $N$ time steps, node $v_i$ collects a block
$U_i^N$ of source symbols -- we refer to the collection of all blocks
$[U_0^N(k)U_1^N(k)...U_M^N(k)]$ collected at time $kN$ ($k\geq 1$) as
a {\em block of snapshots}.  Node $v_i$ then sends a codeword
${X}_{ij}^N$ to node $v_j$.  This codeword depends on a {\em window}
of $K$ previous blocks of source sequences $U_i^N$ observed at node
$v_i$, and of $T$ previously received blocks of channel outputs,
corresponding to noisy versions of the codewords sent by all nodes to
node $v_i$ in the previous $T$ communications steps (corresponding to
$NT$ time steps).

For a block of snapshots observed at time $kN$, at time $(k+W)N$ (that
is, after allowing for a finite but otherwise arbitrary amount of time
to elapse,\footnote{During the time that a block of snapshots spends
within the network, arbitrarily complex coding operations are allowed
within the pipeline: nodes can exchange information, redistribute their
load, and in general perform any form of joint source-channel coding
operations.  The only constraint imposed is that all information
eventually be delivered to destination, within a finite time horizon.}
in which the information injected by all nodes 
reaches $v_0$), an attempt is made to decode at $v_0$.  The decoder produces
an estimate of the block of snapshots $U_0^N(k)U_1^N(k)...U_M^N(k)$ based on
the local observations $U_0^N(k)$, and the previous $W$ blocks of $N$
channel outputs generated by codewords sent to $v_0$ by the other nodes.

Thus, a {\em code} for this network consists of:
\begin{itemize}
\item four integers $N$, $K$, $T$ and $W$;
\item encoding functions at each node
  \[ g_{ij}:\bigotimes_{l=1}^K\mathcal{U}_i^N \times
            \bigotimes_{t=1}^T\bigotimes_{m=0}^M \mathcal{Y}_{mi}^N
            \longrightarrow \mathcal{X}_{ij}^N,
  \]
  for $0 \leq i, j \leq M$.
\item the decoding function at node $v_0$:
  \[ h: \mathcal{U}_0^N \times
        \bigotimes_{w=1}^W\bigotimes_{m=1}^M \mathcal{Y}_{m0}^N
        \longrightarrow \bigotimes_{m=1}^M \hat{\mathcal{U}}_m^N.
  \]
\item the block probability of error:
  \[ P_e^{(N)} = P(U_1^N...U_M^N\neq\hat{U}_1^N...\hat{U}_M^N). \]
\end{itemize}

We say that blocks of snapshots $U_1^N...U_M^N$ can be
{\em reliably communicated} to $v_0$ if there exists a sequence of
codes as above, with $P_e^{(N)}\to 0$ as $N\to\infty$, for some finite
values $K$, $T$ and $W$, all independent of $N$.

With these definitions, we are now ready to state our main theorem.

\begin{theorem}
Let $S$ denote a non-empty subset of node indices that does not contain
node $0$: $S \subseteq \{0...M\}$, $S\neq\emptyset$, $0\in S^c$.  Then,
it is possible to communicate $U_1...U_M$ reliably to $v_0$ if and
only if, for all $S$ as above,
\begin{equation}
 H(U_S|U_{S^c}) \;\;<\;\; \sum_{i\in S,j\in S^c} C_{ij}.
  \label{eq:main}
\end{equation}
\label{thm:main}
\end{theorem}

\subsection{Achievability Proof}

Our coding strategy is based on separate source and channel coding.
We first use capacity attaining channel codes to turn the noisy network
into a network of noiseless links (of capacity $C_{ij}$).  Then, we
use Slepian-Wolf source codes, jointly with a custom designed routing
algorithm, to deliver all this data to destination.  Since the channel
coding aspects of the proof are rather straightforward extensions of
classical point-to-point arguments, in the following we only focus on
the less obvious source coding and routing aspects.

\subsubsection{Mechanics of the Coding Strategy}

Consider a ``noise-free'' version of the problem formulated above: we
still have a complete graph, now with {\em noiseless} links of capacity
$C_{ij}$.  Variables $U_i$ are still observed at each node $v_i$, and
the goal remains to reproduce all of these at $v_0$.  Each node uses
a classical Slepian-Wolf code: there is a source encoder at node $v_i$
that maps a sequence $U_i^N$ to an index from the random binning set
$\{1,2,\dots,2^{NR_i}\}$, thus compressing the block of observations
$U_i^N$ using codes as in~\cite[Thm.\ 14.4.2]{CoverT:91}.  Let
$(R_1...R_M)$ denote the rate allocation to each of the nodes.  To
achieve perfect reconstruction, these bits must be delivered to node
$v_0$.

\begin{itemize}
\item Set $K=T=1$ -- each block of source symbols and each block of
  codewords participates in the encoding process only once.
\item To deliver the bin indices produced by the Slepian-Wolf codes
  to destination, the noise-free network is regarded as a flow
  network~\cite[Ch.\ 26]{CormenLRS:01}.
  Let $\flow(v_i,v_j)$ be a feasible flow in this network, with $M$ sources
  $v_1...v_M$, supply $R_i$ at source $v_i$, and a single sink $v_0$.
  If no such feasible flow exists, the code construction fails.
\item If there is a feasible flow $\flow$ then this $\flow$ uniquely
  determines, at each node $v_i$, the number of bits that need to be
  sent to each of its neighbors -- thus from $\flow$ we derive the
  encoding functions $g_{ij}$ as follows:
  \begin{itemize}
  \item Consider the directed {\em acyclic} graph $G'$ of $G$ induced by
    $\flow$, by taking $V(G') = V(G)$, and
    $E(G')=\{(v_i,v_j)\in E:\flow(v_i,v_j)>0\}$.  Define a permutation
    $\pi:\{0...M\}\to\{0...M\}$, such that
    $[v_{\pi(0)}v_{\pi(1)}...v_{\pi(M)}]$ is a {\em topological sort} of
    the nodes in $G$, as illustrated in Fig.~\ref{fig:topological-sort}.
    \begin{figure}[ht]
    \centerline{\psfig{file=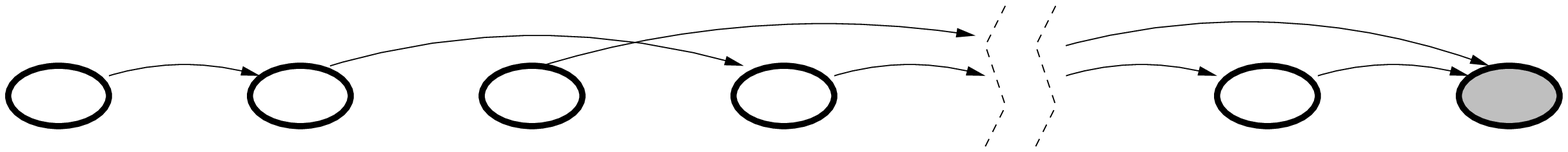,width=16cm,height=2.5cm}}
    \vspace{-5mm}
    \caption{\small A topological sort of the nodes of a directed acyclic
      graph is a linear ordering $v_1...v_M$ such that if $(v_i,v_j)$ is
      an edge, then $i<j$.}
    \label{fig:topological-sort}
    \end{figure}
  \item Consider a block of snapshots
    ${\bf U}(k)=[U_0^N(k)U_1^N(k)...U_M^N(k)]$ captured at time $kN$.
    At time $(k+l)N$ (for $l=0...M$), node $v_{\pi(l)}$ will have received
    all bits with portions of the encodings of ${\bf U}(k)$ generated by
    nodes upstream in the topological order -- thus, together with its own
    encoding of $U^N_{\pi(l)}(k)$, all the bits for ${\bf U}(k)$ up
    to and including node $v_{\pi(l)}$ will be available there, and thus
    can be routed to nodes downstream in the topological order.
  \item Consider now all edges of the form $(v_{\pi(k)},v')$ for which
    $\flow(v_{\pi(k)},v') > 0$:
    \begin{enumerate}
    \item Collect the $m=\sum_{v'} \flow(v',v_{\pi(k)})$ information bits
      sent by the upstream nodes $v'$.
    \item Consider now the set of all downstream nodes $v''$, for which
      $\flow(v_{\pi(k)},v'') > 0$.  Due to flow conservation for $\flow$,
      $\sum_{v''} \flow(v_{\pi(k)},v'')=m+R_{\pi(k)}$, where
      $R_{\pi(k)}$ is the rate allocated to node $v_{\pi(k)}$.
    \item For each $v''$ as above, define $g_{\pi(k)v''}^{(k)}$ to be a
      message such that $|g_{\pi(k)v''}^{(k)}| = \flow(v_{\pi(k)},v')$.
      Partition the $m+R_{\pi(k)}$ available bits according to the values
      of $\flow$, and send them downstream, as illustrated in
      Fig.~\ref{fig:shuffle}.
      \begin{figure}[!ht]
      \centerline{\psfig{file=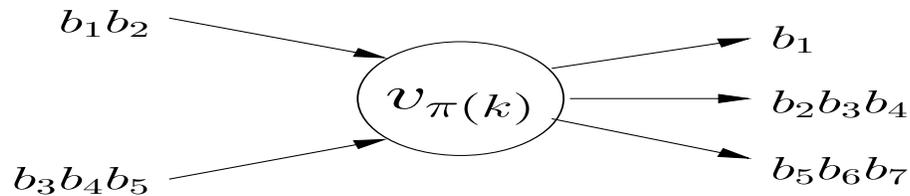,height=2.5cm,width=12cm}}
      \vspace{-3mm}
      \caption{\small To illustrate the operations performed at each node.
        In this example, five bits come into node $v_{\pi(k)}$ from 
        neighbouring nodes, two on the top link and three on the bottom
        link.  The information bits from other nodes come in the form of
        noisy codewords -- they need to be decoded from the received channel
        outputs.  Now, because flow conservation holds for $\flow$, we
        know that the aggregate capacity of the three output links will
        be at least five bits plus some local bits (the encoding of a
        block of local observations $U^N_{\pi(k)}$, denoted by $b_6$ and
        $b_7$ here).  So at this point we split those bits in a way such
        that the individual capacity constraints of the output links are
        not violated, and then they are sent on their way to $v_0$.}
      \label{fig:shuffle}
      \end{figure}
    \end{enumerate}
  \end{itemize}
\item To decode, at time $(k+M)N$, node $v_0$ does the following:
  \begin{itemize}
  \item Decode all channel outputs received at time $(k+M-1)N$, to recover
    the bits sent by each 1-hop neighbor of the sink.
  \item Reassemble the set of bin indices from the segments received
    from each neighbor.
  \item Perform typical set decoding (as in~\cite[pg.\ 411]{CoverT:91}),
    to recover the block of snapshot $[U_1^N(k)...U_M^N(k)]$.
  \end{itemize}
\end{itemize}
An important observation is that, in this setup, network coding (in the
sense of~\cite{AhlswedeCLY:00}) is not needed.  This is because we have
a case of $M$ sources and a single sink interested in collecting all
messages, a case for which it was shown in~\cite{LehmanL:04} that routing
alone suffices.

Our next task is to find conditions under which this coding strategy
results in $P_e^{(N)}\to 0$ as $N\to\infty$.

\subsubsection{Analysis of the Probability of Error}

The coding strategy proposed above hinges on two main elements:
\begin{itemize}
\item Slepian-Wolf codes: in this case, we know that provided the rate
  vector $(R_1...R_M)$ is such that, for all partitions $S$ of $\{0...M\}$,
  $S\neq\emptyset$, $0\in S^c$,
  \begin{equation}
  \sum_{i\in S} R_i > H(U_S|U_{S^c}),
  \label{eq:achievability1}
  \end{equation}
  then there exist Slepian-Wolf codes with arbitrarily low probability of
  error~\cite[Ch.\ 14.4]{CoverT:91}.
\item Network flows: from elementary flow concepts we know that if a
  flow $\flow$ is feasible in a network $G$, then for all
  $S\subseteq\{0...M\}$, $S\neq\emptyset$, $0\in S^c$,
  \begin{eqnarray}
  \sum_{i\in S} R_i
     & \stackrel{(a)}{=} & \sum_{i\in S,j\in V} \flow(v_i,v_j) \nonumber \\
     & \stackrel{(b)}{=} & \sum_{i\in S,j\in S^c} \flow(v_i,v_j) \nonumber \\
     & \stackrel{(c)}{\leq} & \sum_{i\in S,j\in S^c} C_{ij},
     \label{eq:achievability2}
  \end{eqnarray}
  where $(a)$ and $(b)$ follow from the flow conservation properties of a
  feasible flow (all the flow injected by the sources has to go somewhere
  in the network, and in particular all of it has to go across a network
  cut with the destination on the other side); and $(c)$ follows from the
  fact that in any flow network, the capacity of any cut is an upper bound
  to the value of any flow.
\end{itemize}
Thus, from~(\ref{eq:achievability1}) and~(\ref{eq:achievability2}), we
conclude that if, for all partitions $S$ as above, we have that
\begin{equation}
  H(U_S|U_{S^c}) < \sum_{i\in S,j\in S^c} C_{ij},
\end{equation}
then $P_e^{(N)}\to 0$ as $N\to\infty$.

\subsection{Converse Proof}

The converse proof is fairly long and tedious, but by virtue of being
based on Fano's inequality and standard information-theoretic arguments,
it is relatively straightforward -- therefore, we omit it here and
provide the technical details in Appendix~\ref{app:proof-converse-mcoop}.
At this point however, we would like to sketch out an informal argument
on why this converse should hold.

\begin{figure}[!ht]
\centerline{\psfig{file=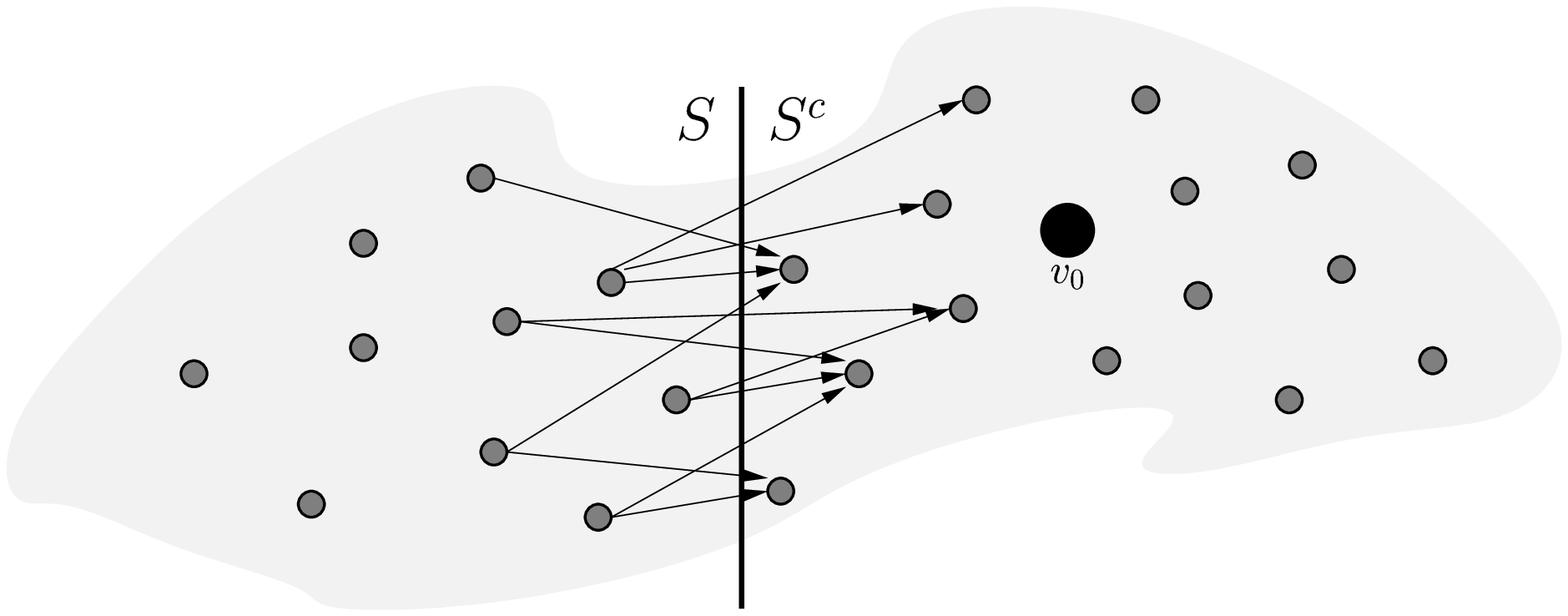,height=3cm,width=15cm}}
\caption{\small An artificial two-terminal system: all sources in $S$
  are treated as a supersource, connected to a supersink made of all
  the sinks in $S^c$ by an array of DMCs (those going across the cut).
  Intuitively, any necessary condition for this system should also be
  necessary for our system (although this requires a formal statement
  and proof).  The interesting statement thus is to show that the
  set of all conditions obtained in this form (by considering all
  possible cuts) is also sufficient.}
\label{fig:oracle}
\end{figure}

Consider an arbitrary network partition $S$ of $\{0...M\}$,
$S\neq\emptyset$, $0\in S^c$.  For each such partition we define a
two-terminal system, with a ``supersource'' that has access to the
whole vector of observations $U_1...U_M$, and a ``supersink'' that
has access only to $U_{S^c}$.  The supersource and supersink are
connected by an array of parallel DMCs: if $i\in S$ and $j\in S^c$,
then $(\mathcal{X}_{ij},p_{ij}(y|x),\mathcal{Y}_{ij})$ from the
network is one of the channels in the array.  This is illustrated
in Fig.~\ref{fig:oracle}.

Clearly, $H(U_S|U_{S^c}) < \sum_{i\in S,j\in S^c} C_{ij}$ is an outer
bound for this two-terminal system (follows directly from the source/channel
separation theorem,~\cite[Sec.\ 8.13]{CoverT:91}).  And intuitively,
it is also clear that any outer bound for this two-terminal system
provides necessary conditions for reliable communication to be possible
in our network.  Thus, by considering all possible partitions $(S,S^c)$
as above, we obtain a set of necessary conditions matching those of the
achievability result.\footnote{We thank our Reviewer B, for suggesting
this simple and very clear interpretation for the converse.}

We would also like to highlight that, because of the correlation between
sources, a simple max-flow/min-cut bounding argument as suggested
in~\cite[Section 14.10]{CoverT:91}) is not sufficient to establish the
source-channel separation result we seek -- proving said result requires
all the steps of a typical converse.

A formal proof for this converse is provided in
Appendix~\ref{app:proof-converse-mcoop}.

\subsection{Special Cases}

\subsubsection{A Network with Three Nodes}
\label{sec:three-nodes}

\begin{figure}[!ht]
\centerline{\psfig{file=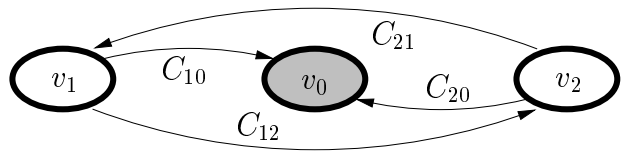,height=2cm}}
\caption{\small A network with three nodes.}
\label{fig:three-nodes}
\end{figure}

To provide an illustration of the meaning of Theorem~\ref{thm:main}, and
of the optimality of the flow-based solution, we specialize
Theorem~\ref{thm:main} to the case of a network with three nodes.  In
this case, those conditions become:
\begin{eqnarray}
H(U_1|U_2U_0) & < & C_{10} + C_{12} \label{eq:3nodes-1} \\
H(U_2|U_1U_0) & < & C_{20} + C_{21} \label{eq:3nodes-2} \\
H(U_1U_2|U_0) & < & C_{10} + C_{20} \label{eq:3nodes-3}.
\end{eqnarray}
A network with three nodes as considered here is illustrated in
Fig.~\ref{fig:three-nodes}.

\begin{figure}[ht]
\centerline{\psfig{file=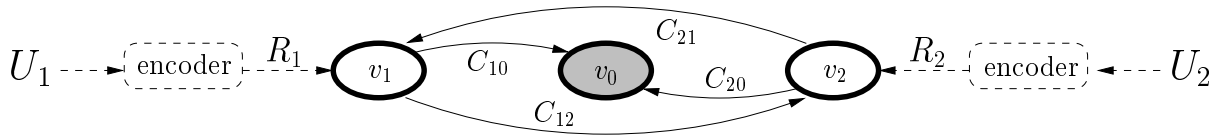,height=1.8cm}}
\caption{\small A flow network with three nodes, supplies $R_1$ and
  $R_2$ and nodes $v_1$ and $v_2$, and a sink $v_0$.}
\label{fig:three-nodes-flownetwork}
\end{figure}

Next, we regard the network in Fig.~\ref{fig:three-nodes} as a
{\em flow} network~\cite[Ch.\ 26]{CormenLRS:01}: a flow network with
two sources ($v_1$ and $v_2$) and a single sink ($v_0$).  Encodings
of $U_1$ injected at source $v_1$ at rate $R_1$, and of $U_2$ injected
at $v_2$ at rate $R_2$, are the ``objects'' that flow in this network
and are to be delivered to the sink $v_0$.  This is illustrated in
Fig.~\ref{fig:three-nodes-flownetwork}.

In the simple flow network of Fig.~\ref{fig:three-nodes-flownetwork},
any feasible flow $\flow$ must satisfy some {\em conservation} equations:
\[\begin{array}{rclcl}
  R_1 & = & \flow(v_1,v_0)+\flow(v_1,v_2), \\
  R_2 & = & \flow(v_2,v_0)+\flow(v_2,v_1), \\
  R_1+R_2 & = & \flow(v_1,v_0)+\flow(v_1,v_2)+\flow(v_2,v_0)+\flow(v_2,v_1)
          & = & \flow(v_1,v_0)+\flow(v_2,v_0),
\end{array}\]
where the last equality follows from the fact that flow conservation
holds: the total amount of flow injected ($R_1+R_2$) must equal the total
amount of flow received by the sink
($\flow(v_1,v_0)+\flow(v_2,v_0)$)~\cite{CormenLRS:01}.  Similarly, any
feasible flow must also satisfy all {\em capacity} constraints:
\[\begin{array}{rcl}
  \flow(v_1,v_0)+\flow(v_1,v_2) & \leq & C_{10}+C_{12}, \\
  \flow(v_2,v_0)+\flow(v_2,v_1) & \leq & C_{20}+C_{21}, \\
  \flow(v_1,v_0)+\flow(v_2,v_0) & \leq & C_{10}+C_{20}.
\end{array}\]
Combining these last two sets of constraints, and the conditions from
the Slepian-Wolf theorem on feasible $(R_1,R_2)$ pairs, we immediately
get
\[\begin{array}{rcccl}
  H(U_1|U_2U_0) & < & R_1 & \leq & C_{10}+C_{12}, \\
  H(U_2|U_1U_0) & < & R_2 & \leq & C_{20}+C_{21}, \\
  H(U_1U_2|U_0) & < & R_1+R_2 & \leq & C_{10}+C_{20}.
\end{array}\]

\begin{figure}[ht]
\centerline{\psfig{file=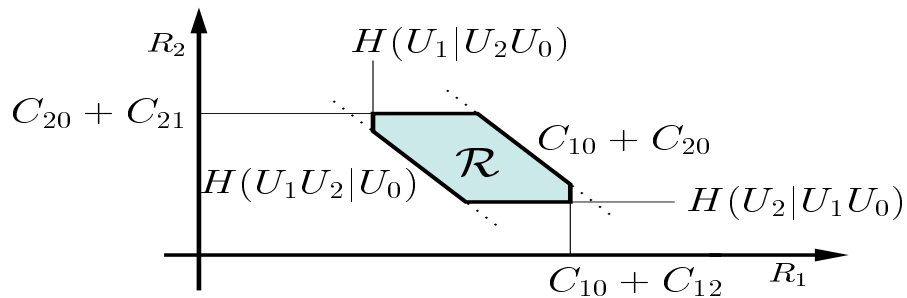,width=12cm,height=4cm}}
\caption{The polytope $\mathcal{R}$ of admissible rates.}
\label{fig:polytope}
\end{figure}

It is interesting to observe in this argument that the region of
achievable rates forms a convex polytope, in which three of its
faces come from the Slepian-Wolf conditions, and three come from
the capacity constraints.  This polytope is illustrated in
Fig.~\ref{fig:polytope}.
This polytope plays a central role in our analysis: reliable
communication is possible {\em if and only if} $\mathcal{R}\neq\emptyset$.
Thus, the view of ``information as a flow'' in this class of networks
is complete.

\subsubsection{No Cooperation and No Side Information at $v_0$}

\begin{figure}[!ht]
\centerline{\psfig{file=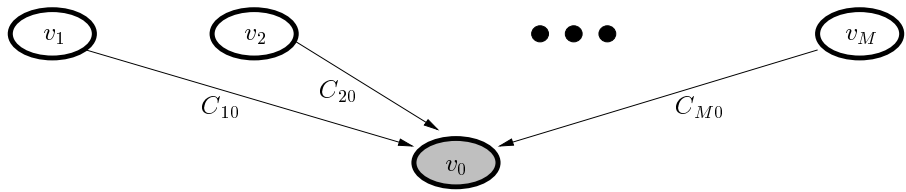,width=12cm,height=2.5cm}}
\caption{\small M non-cooperating nodes.}
\label{fig:m-nodes}
\end{figure}

We consider now the special case of $M$ {\em non-}cooperating nodes and
one sink, as illustrated in Fig.~\ref{fig:m-nodes}.
Necessary and sufficient conditions for reliable communication
under this scenario follow naturally from our main theorem by setting
$C_{ij}=0$ for all $j\neq 0$, and $|\mathcal{U}_0 | = 1$.

\begin{corollary}
\label{cor:indep}
The sources $U_1, U_2,\dots, U_M$ can be communicated reliably over an
array of independent channels of capacity $C_{i0}$, $i=1\dots M$, if and
only if
\[
  H(U_S|U_{S^c})<\sum_{i\in S}C_{i0},
\]
for all subsets $S\subseteq\{1,2,\dots,M\}$, $S\neq\emptyset$.
\end{corollary}

\begin{figure}[ht]
\centerline{\psfig{width=8cm,height=5cm,file=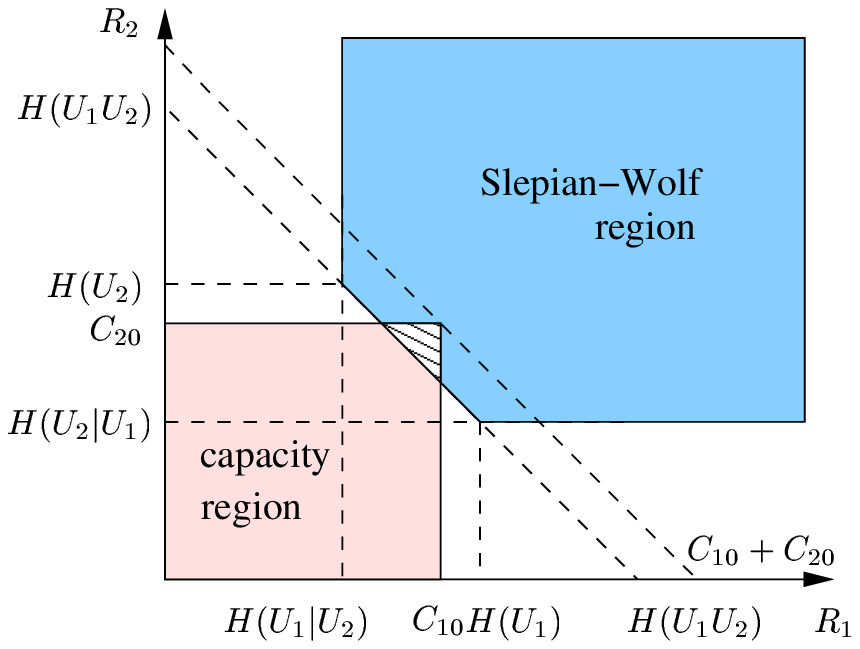}
            \psfig{width=8cm,height=5cm,file=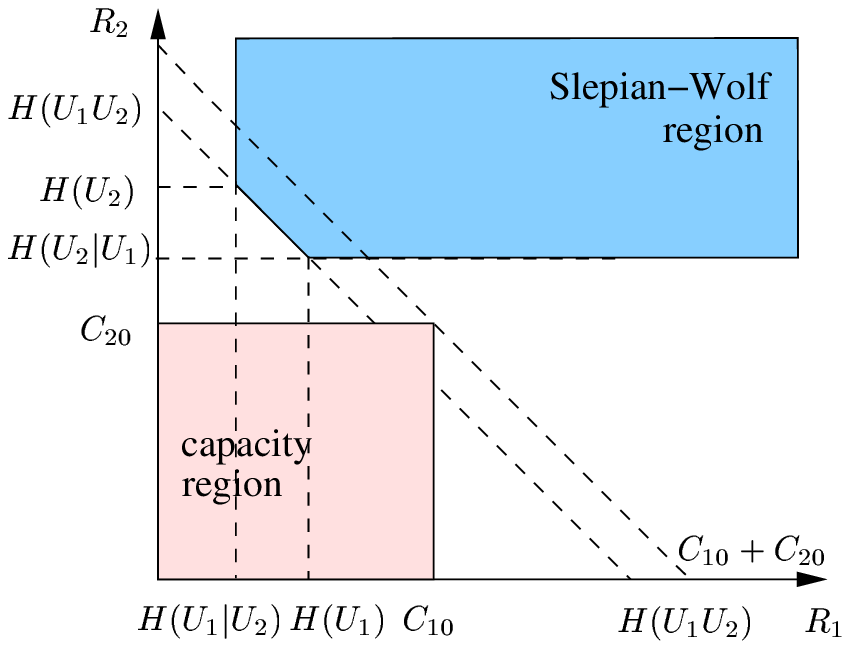}}
\caption[Relationship between the Slepian-Wolf region and the capacity
  region for two independent channels.]
  {Relationship between the Slepian-Wolf region and the capacity
  region for two independent channels. In the left figure, as
  $H(U_1|U_2)<C_{10}$ and $H(U_2|U_1)<C_{20}$ the two regions intersect
  and therefore reliable communication is possible. The figure on the
  right shows the case in which $H(U_2|U_1)>C_{20}$ and there is no
  intersection between the two regions.}
\label{fig:SWMAC1}
\end{figure}

An illustration of this corollary for two sources $U_1$ and $U_2$ is
shown in Fig.~\ref{fig:SWMAC1}.
When we have two independent channels with capacities $C_{10}$ and
$C_{20}$, the capacity region becomes a rectangle with side lengths
$C_{10}$ and $C_{20}$~\cite[Chapter~14.3]{CoverT:91}.
Also shown is the Slepian-Wolf region of achievable rates for separate
encoding of correlated sources.
Clearly, $H(U_1U_2)<C_{10}+C_{20}$ is a necessary
condition for reliable communication as a consequence of Shannon's joint
source and channel coding theorem for point-to-point communication.
Assuming that this is the case, consider now the following possibilities:
\begin{itemize}
\item $H(U_1)<C_{10}$ and $H(U_2)<C_{20}$.  The Slepian-Wolf region and the
  capacity region
  intersect, so any point $(R_1,R_2)$ in this intersection makes reliable
  communication possible.  Alternatively, we can argue that reliable
  transmission of $U_1$ and $U_2$ is possible even with independent decoders,
  therefore a joint decoder will also achieve an error-free reconstruction
  of the source.
\item $H(U_1)>C_{10}$ and $H(U_2)>C_{20}$.  Since $H(U_1U_2)<C_{10}+C_{20}$
  there is always at least one point of intersection between the Slepian-Wolf
  region and the capacity region, so reliable communication is possible.
\item $H(U_1)<C_{10}$ and $H(U_2)>C_{20}$ (or vice versa).  If
  $H(U_2|U_1)<C_{20}$ (or if $H(U_1|U_2)<C_{10}$) then the two regions will
  intersect.  On the other hand, if $H(U_2|U_1)>C_{20}$ (or if
  $H(U_1|U_2)>C_{10}$), then there are no intersection
  points, but it is not immediately clear whether reliable communication
  is possible or not (see Fig. \ref{fig:SWMAC1}), since examples are known
  in which the intersection between the capacity region of the multiple
  access channel and the Slepian-Wolf region of the correlated sources
  is empty and still reliable communication is possible~\cite{CoverGS:80}.
\end{itemize}
Corollary~\ref{cor:indep} gives a definite answer to this last question:
in the special case of correlated sources and independent channels an
intersection between the capacity region and the Slepian-Wolf rate regions
is not only sufficient, but also a necessary condition for reliable
communication to be possible---in this case, separation holds.

\subsubsection{Arrays of Gaussian Channels}

We should also mention that Theorem~\ref{thm:main} applies to other
channel models that are relevant in practice, for instance Gaussian channels
with orthogonal multiple access.  For simplicity, we illustrate 
this issue  in the context of
Corollary~\ref{cor:indep}. The capacity of the Gaussian
multiple access channel with $M$ independent sources is given by
\[
  \sum_{i\in S} R_i
    \leq \frac{1}{2}\log\left(1+\frac{\sum_{i\in S}P_i}{\sigma^2}\right),
\]
for all $S\subseteq\{1...M\}$, $S\neq\emptyset$, and where $\sigma^2$ and
$P_i$ are the noise power and the power of the $i$-th user
respectively~\cite[pp.\ 378-379]{CoverT:91}.  If we use orthogonal accessing
(e.g.~TDMA), and assign different time slots to each of the transmitters,
then the Gaussian multiple access channel is reduced to an array of $M$
independent single-user Gaussian channels each with capacity
\[
  C_{i0} =
  \tau_{i0}\cdot\frac{1}{2}\log\bigg(1+\frac{P_{i0}}{\sigma^2\tau_{i0}}\bigg),
  \qquad 1\le i\le M,
\]
where $\tau_{i0}$ is the time fraction allocated to source user $i$ to
communicate with the data collector node $v_0$, and $P_{i0}$ is the
corresponding power allocation.

Applying Theorem~\ref{thm:main}, we obtain the reachback capacity
of the Gaussian channel with orthogonal accessing.\footnote{The
generalization of Theorem~\ref{thm:main} for channels with real-valued
output alphabets can be easily obtained using the techniques
in~\cite[Sec.\ 9.2 \& Ch.\ 10]{CoverT:91}.}  Then, reliable
communication is possible if and only if
\[
   H(U_S|U_{S^c}) \leq
     \sum_{i\in S}\frac{\tau_{i0}}{2}
                  \log\bigg(1+\frac{P_{i0}}{\sigma^2\tau_{i0}}\bigg),
\]
for all subsets $S\subseteq\{1,2,\dots,M\}$, $S\neq\emptyset$.

\section{Practical/Engineering Implications of Theorem~\ref{thm:main}}
\label{sec:protocol-stack}

\subsection{An Information Theoretically Optimal Protocol Stack}

\begin{figure}[!ht]
\centerline{\psfig{file=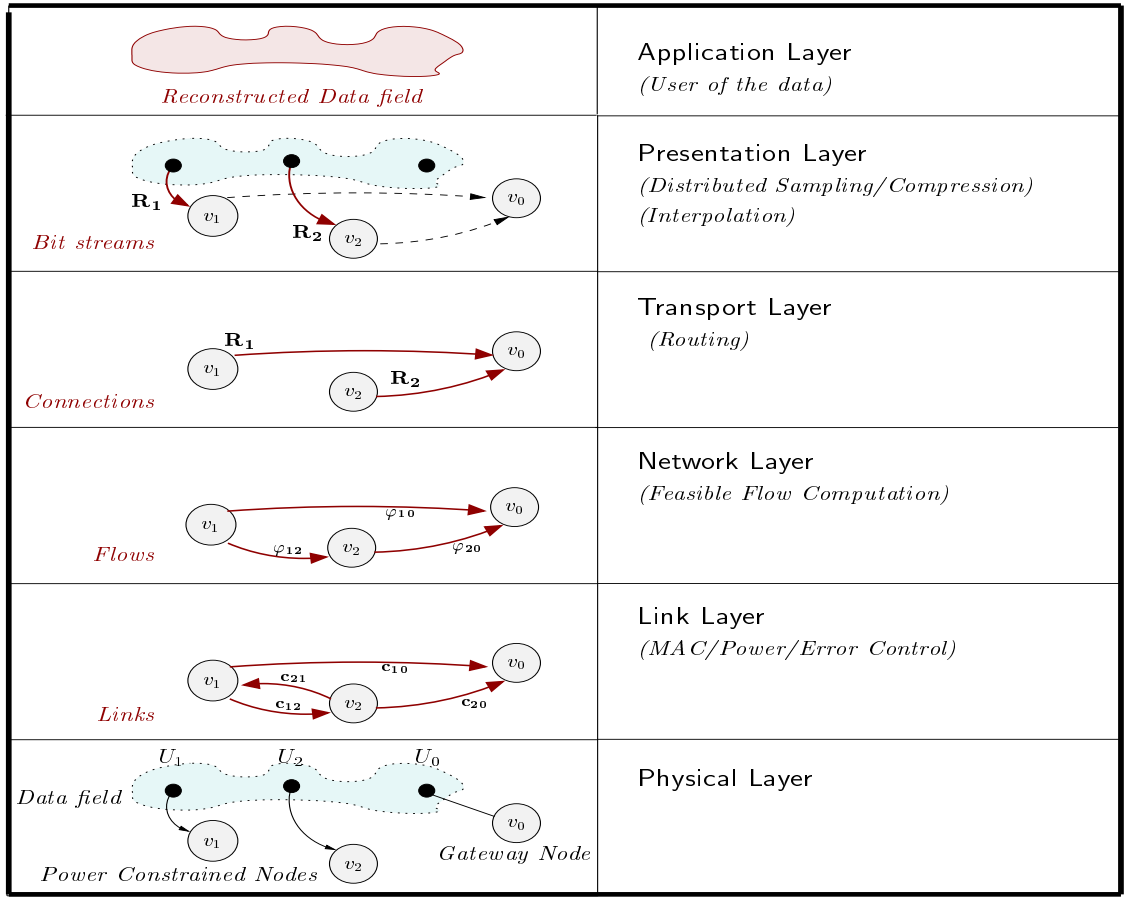,width=15cm,height=12cm}}
\caption{\small Abstractions that follow from the achievability proof,
  illustrated here for three nodes.  At the physical layer there are
  nodes with power constraints, a data field of which these nodes collect
  samples in space and time, and a gateway node that will deliver all
  this data to destination.  On top of this physical substrate, we
  construct a sequence of abstractions: noiseless point-to-point links
  of a given capacity (the {\em Link Layer}); a flow network (the
  {\em Network Layer}); a set of connections (the {\em Transport Layer});
  and a set of distributed signal processing algorithms for sampling,
  compression and interpolation of the space/time continuous process
  (the {\em Presentation Layer}).  In the end, an approximate
  representation of the underlying data field is delivered to
  applications.}
\label{fig:layers-3nodes}
\end{figure}

We believe that the fact that in networks of point-to-point noisy
links with one sink
 Shannon information has the exact same properties of classical
network flows is of particular {\em practical} relevance.  This is
so because there is a rich {\em algorithmic} theory associated with
it, which allows us to cast standard information theoretic problems
into the language of flows and optimization.  Perhaps most relevant
among these is is the optimality of implementing codes using a
{\em layered} protocol stack, as illustrated in
Fig.~\ref{fig:layers-3nodes}.
 
As discussed in the Introduction, the decision to turn a wireless
network into a network of point-to-point links is an arbitrary one.
But, due to complexity and/or economic considerations, this arbitrary
decision is one made very often, and thus we believe it is of great
practical interest to understand what are appropriate design criteria
for such networks.  And our Theorem~\ref{thm:main} offers valuable
insights in this regard -- {\em if} we decide to define a link-layer
based on a MAC protocol that deals with interference by suppressing
it, {\em then all remaining layers in Fig.~\ref{fig:layers-3nodes}
follow from the achievability proof of Theorem~\ref{thm:main}.}  We
see therefore that indeed, in this class of networks,
Fig.~\ref{fig:layers-3nodes} provides a set of abstractions analogous
to those of Fig.~\ref{fig:shannon-pt2pt} for classical two-terminal
systems.

\subsection{Algorithmic/Computational Issues}
\label{sec:algorithmic-issues}

As an illustration of the benefits of the ``information as flow''
interpretation for our results, in this subsection we outline some
initial results on an optimal routing problem.  This topic however
will be developed in full depth elsewhere.

\subsubsection{Optimization Aspects of Protocol Design}

A natural question that follows from our previous developments is
one of {\em optimization}: given a non-empty feasibility polytope
$\mathcal{R}$, we have the freedom of choosing among multiple
assignments of values to flow variables, and thus it is only natural
to ask if there is an optimal flow.  To this end, we define a cost
function $\kappa$ as follows:
\[
  \kappa(\flow) = \sum_{(v_i,v_j)\in E} c(v_i,v_j)\cdot\flow(v_i,v_j),
\]
where $c(v_i,v_j)$ is a constant that, multiplied by the total number
of bits $\flow(v_i,v_j)$ that a flow $\flow$ assigns to an edge
$(v_i,v_j)$, determines the cost of sending all that information over
the channel $(\mathcal{X}_{ij},p_{ij}(y|x),\mathcal{Y}_{ij})$.  The
resulting optimization problem is shown in Fig.~\ref{fig:lp-optrouting}.

\begin{figure}[ht]
\begin{center}
\fbox{\begin{minipage}{11.7cm}
min \hspace{5mm} $\sum_{(v_i,v_j)\in E}\;\;c(v_i,v_j)\cdot\flow(v_i,v_j)$ \\
$\;$ subject to: \vspace{-1mm}
\[\begin{array}{lll}
 & \mbox{\tiny\sl Standard flow constraints (capacity / skew symmetry / flow conservation)} \\
 & \flow(v_i,v_j) \leq C_{ij}, & 0\leq i,j\leq M. \\
 & \flow(v_i,v_j) = -\flow(v_j,v_i), & 0\leq i,j\leq M. \\
 & \sum_{v\in V} \flow(v_i,v) = 0, & 1\leq i\leq M. \\
 & \mbox{\tiny\sl Rate admissibility constraints} \\
 & H(U_S|U_{S^c}) < \sum_{i\in S} \flow(s,v_i)
                  \leq \sum_{i\in S,j\in S^c} C_{ij},
   & S\subseteq\{1...M\}, S\neq\emptyset. \\
 & \flow(s,v_i) = R_i, & 1\leq i\leq M.
\end{array}\]
\end{minipage}}\end{center}
\vspace{-1mm}
\caption{\small Linear programming formulation for the assignment of
  values to flow variables (observe the introduction of a ``supersource''
  $s$, which supplies $R_i$ units of flow to $v_i$).  A solution to this
  problem provides optimal routes (those with positive flow assignment)
  and loads on each link.  Note as well that, by choosing $c(v_i,v_j)=0$
  for all $(v_i,v_j)\in E$, this LP is solvable if and only if
  $\mathcal{R}\neq\emptyset$ -- that is, the decision problem for reliable
  communication (i.e., for whether a given load $p(U_0U_1...U_M)$ can be
  carried over a given network $G$) admits a linear programming formulation
  too.}
\label{fig:lp-optrouting}
\end{figure}

The choice of a linear cost model in this setup can be justified based
on a number of reasons.  First of all, linearity is a very natural
assumption: in simple language, it says that it costs twice as much to
double the amount of information sent on any channel.  For example, we
could take $c(v_i,v_j)$ to be the {\em minimum energy per information
bit} required for reliable communication over the DMC from $v_i$ to
$v_j$~\cite{Verdu:02}, and then $\kappa(\flow)$ would give us the sum of
the energy consumed by all nodes when transporting data as dictated
by a particular flow $\flow$.  Specifically in the context of routing
problems, another important consideration is that the main drawback
often cited for solving optimal routing problems based on network flow
formulations is given by the fact that cost functions such as $\kappa$
only optimize {\em average} levels of link traffic, ignoring other
traffic statistics~\cite[pg.\ 436]{BertsekasG:92}.  But this is not
at all an issue here, since the values of flow variables (i.e.,
Shannon information) are already average quantities themselves.

\subsubsection{A Routing Example}

\begin{figure}[ht]
\centerline{\psfig{file=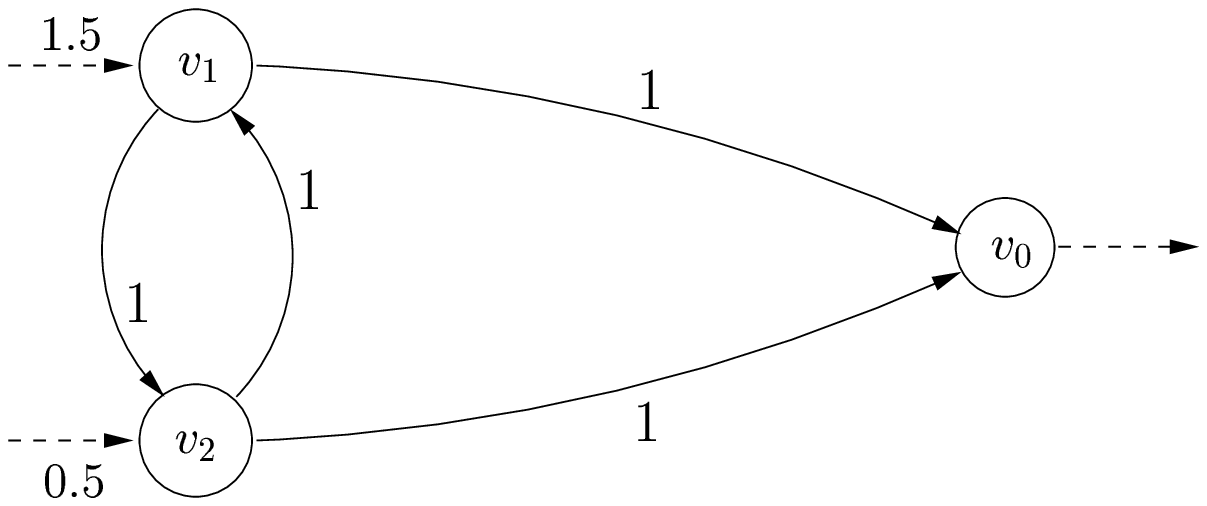,width=5cm,height=3cm}
            \psfig{file=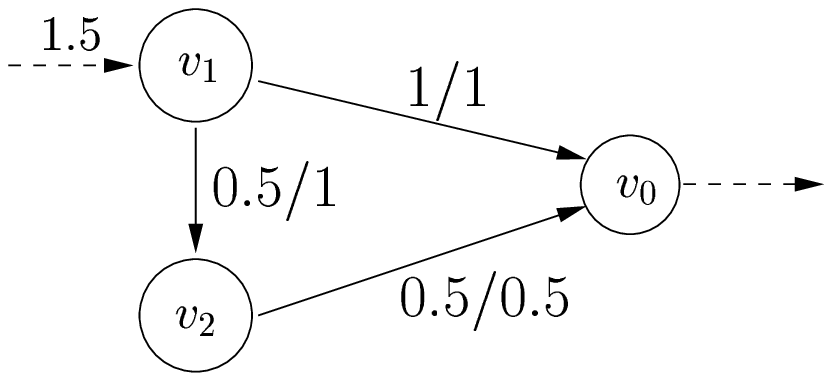,width=5cm,height=3cm}
            \psfig{file=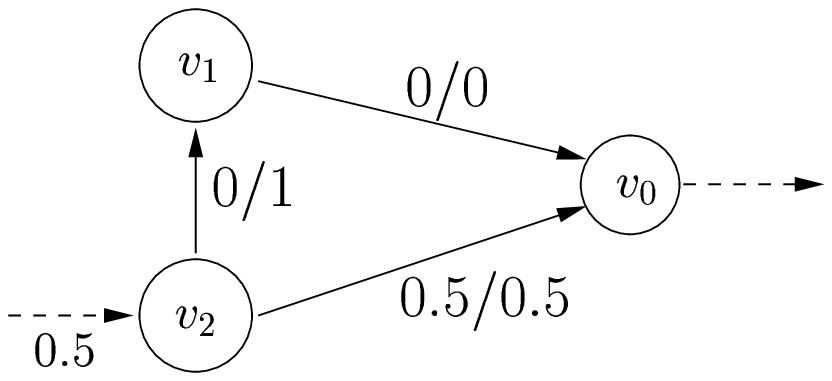,width=5cm,height=3cm}}
\vspace{-2mm}
\caption{To illustrate a solvable problem that cannot be solved using trees.
  Left: a flow network; middle/right: the decomposition of a feasible
  flow into two single flows, showing how much of the flow injected at each
  source is sent over which link ($x/c$ next to an edge means that the
  edge carries $x$ units of flow, and has capacity $c$).}
\label{fig:trees-stink-1}
\end{figure}

As one example of the usefulness of the LP formulation in
Fig.~\ref{fig:lp-optrouting}, we consider next the problem of designing
efficient mechanisms for data aggregation, as motivated
in~\cite{IntanagonwiwatGEHS:03}.  There has been a fair amount of work
reported in the networking literature, on the design and performance
analysis of {\em tree} structures for aggregation---for example, the
work of Goel and Estrin on the construction of trees that perform well
simultaneously under multiple concave costs~\cite{GoelE:03}.  Based on
our LP formulation, we construct two examples which show the extent to
which trees could give rise to suboptimalities, as opposed to other
topological structures.  And we start by showing an example in which,
although $\mathcal{R}\neq\emptyset$, there are no feasible trees.  This
case is illustrated in Fig.~\ref{fig:trees-stink-1}.

As illustrated in Fig.~\ref{fig:trees-stink-1}, a solution to the
transport problem exists.  However, it is easy to check that if we
constrain data to flow along trees, none of the three possible trees
($\{(v_1,v_0);(v_2,v_0)\}$, or $\{(v_1,v_2);(v_2,v_0)\}$, or
$\{(v_2,v_1);(v_1,v_0)\}$) are feasible: in all cases, there is one
link for which the capacity constraint is violated.

\begin{figure}[ht]
\centerline{\psfig{file=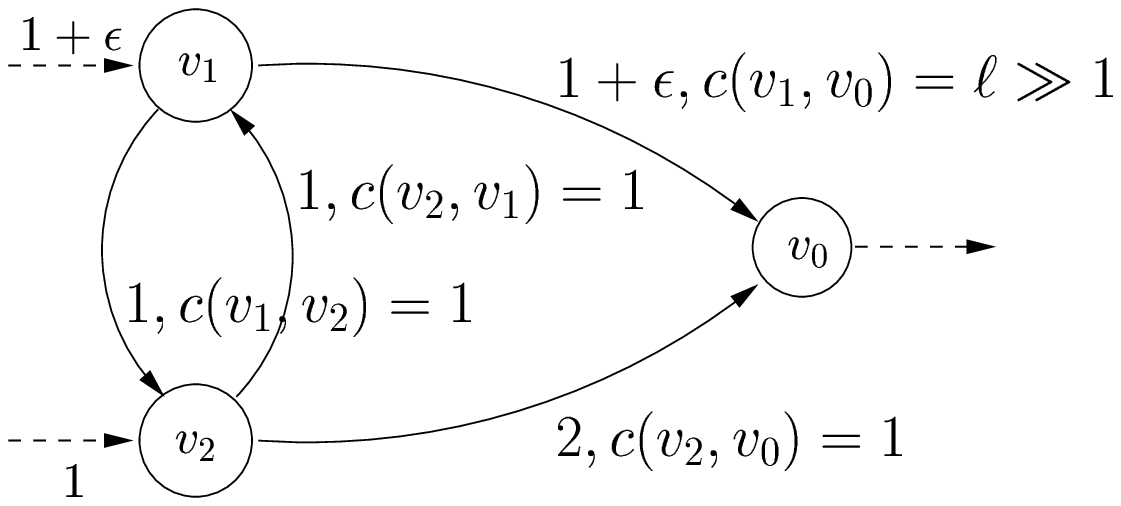,width=5cm,height=3cm}
            \psfig{file=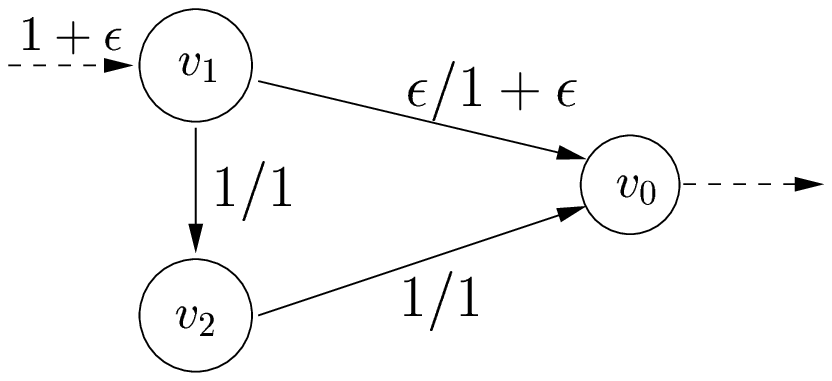,width=5cm,height=3cm}
            \psfig{file=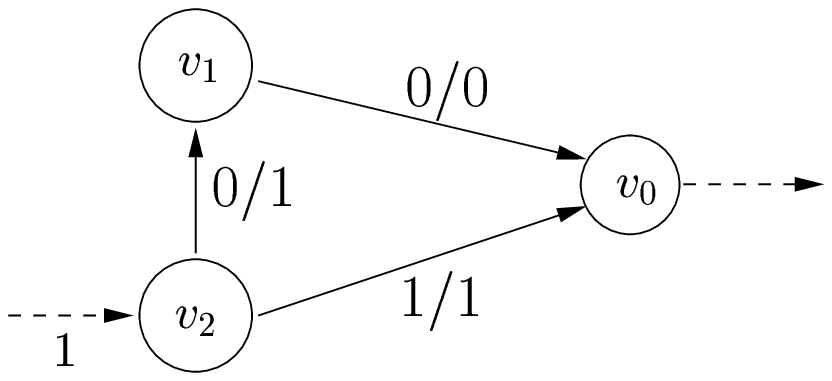,width=5cm,height=3cm}}
\vspace{-2mm}
\caption{To illustrate a problem in which trees are very expensive.
  Left: a flow network with costs; right: an optimal solution to the
  linear program in Fig.~\ref{fig:lp-optrouting}.  Such a case could
  arise, e.g., in a situation where there is heavy interference in
  the direct path from $v_1$ to $v_0$.}
\label{fig:trees-stink-2}
\end{figure}

Next we consider a case where feasible trees exist, but the lowest
cost of any tree differs from the optimal cost by an arbitrarily large
factor.  This case is illustrated in Fig.~\ref{fig:trees-stink-2}.

In this case, there exists only one feasible tree:
$\{(v_1,v_0);(v_2,v_0)\}$, with cost $\ell(1+\epsilon)+1$.  However,
because of the ``expensive'' link $(v_1,v_0)$ along which the tree is
forced to send all its data, the cost is significantly increased:
by splitting the encoding of $U_1$
as illustrated in Fig.~\ref{fig:trees-stink-2}, the cost incurred into
by this structure would be $\epsilon\ell+3$.  Hence, we see that in
this case, the cost of the best feasible tree is
$\frac{\ell(1+\epsilon)+1}{\epsilon\ell+3}$ times larger than that
of an optimal solution allowing splits.  And this
``overpayment factor'' could be significant: when $\ell$ is large,
this is $\approx 1+\frac 1 \epsilon$, and it grows unbound for
small $\epsilon$.

Note as well that any time that a network is operated close to capacity,
it will be necessary to split flows.  And that is a situation likely to
be encountered often in power-constrained networks, since minimum energy
designs will necessarily result in links being allocated the least amount
of power needed to carry a given traffic load.  Thus, we see that these
examples above are {\em not} pathological cases of limited practical
interest, but instead, they are good representatives of situations likely
to be encountered often in practice.

\subsection{Suboptimality of Correlated Codes for Orthogonal Channels}

The key ingredient of the achievability proof presented by Cover,
El Gamal and Salehi for the multiple access channel with correlated
sources is the generation of random codes, whose codewords $X_i^N$ are
statistically dependent on the source sequences $U_i^N$~\cite{CoverGS:80}.
This property, which is achieved by drawing the codewords according to
$\prod_{j=1}^{N}p(x_{ij}|u_{ij})$ with $u_{ij}$ and $x_{ij}$ denoting
the $j$-th element of $U_i^N$ and $X_i^N$, respectively, implies that
$U_i^N$ and $X_i^N$ are jointly typical with high probability.  Since
the source sequences $U_1^N$ and $U_2^N$ are correlated, the codewords
$X_1^N(U_1^N)$ and $X_2^N(U_2^N)$ are also correlated, and so we speak
of {\it correlated codes}.  This class of random codes, which is treated
in more general terms in~\cite{AhlswedeH:83}, can be viewed as joint
source and channel codes that preserve the given correlation structure
of the source sequences, based upon which the decoder can lower the
probability of error.

The class of correlated codes is of interest to us because of two
main reasons:
\begin{itemize}
\item From a practical point of view, correlated codes have a very
  strong appeal: sensor nodes with limited processing capabilities may
  be forced to use very simple codes that do not eliminate correlations
  between measurements prior to transmission~\cite{BarrosTL:03} (e.g.,
  a simple scalar quantizer and simple BPSK modulation).
\item From a theoretical point of view, since these codes yield the
  largest known admissibility region for the problem of communicating
  distributed sources over multiple-access channels, it would be interesting
  to know how these codes fare in our context, where we know separate
  source and channel coding to achieve optimality.
\end{itemize}
Thus, specializing the achievability proof of~\cite{CoverGS:80} to the
case of $M$ independent channels, we get the following result.

\begin{corollary}[From Theorem 1 of~\cite{CoverGS:80}]
\label{cor:Machievable}
A set of correlated sources $[U_1U_2...U_M]$ can be communicated
reliably over independent channels
$(\mathcal{X}_1,p(y_1|x_1),\mathcal{Y}_1)\dots
(\mathcal{X}_M,p(y_M|x_M),\mathcal{Y}_M)$ to a sink $v_0$, if
\[
   H(U_S|U_{S^c})<\sum_{i\in S} I(X_i;Y_0|U_{S^c}),
\]
for all subsets $S\subseteq\{1,2,\dots,M\}$, $S\neq\emptyset$.
\end{corollary}
\begin{proof}
This result can be obtained from the $M$-source version of the main theorem
in ~\cite{CoverGS:80}, by specializing it to a multiple access channel with
conditional probability distribution
\[ p(y|x_1x_2...x_M)
     = p(y_1y_2\dots y_M|x_1x_2\dots x_M) = \prod_{i=1}^Mp(y_i|x_i).
\]
\end{proof}

Part of the reason why we feel this is an interesting result is that the
main theorem in~\cite{CoverGS:80} does {\em not} immediately specialize
to Corollary~\ref{cor:indep}: whereas the achievability results do
coincide,~\cite{CoverGS:80} does not provide a converse.  To illustrate
this point better, we focus now on the case of $M=2$:
\begin{itemize}
\item In general, we have that
  $I(X_1X_2;Y_1Y_2) \leq I(X_1;Y_1)+I(X_2;Y_2)$, for any 
  $p(u_1u_2x_1x_2)p(y_1|x_1)p(y_2|x_2)$; but for this upper bound on
  the sum-rate to be achieved, we must take
  $p(u_1u_2x_1x_2) = p(u_1u_2)p(x_1)p(x_2)$ -- that is, the codewords
  must be drawn independently of the source.  And for this special case,
  our Theorem~\ref{thm:main} does provide a converse.
\item As argued earlier, due to practical considerations it may not be
  feasible to remove correlations in the source before choosing channel
  codewords, in which case we face a situation where correlated codes
  are used, despite their obvious suboptimality.  In this case, it is
  of interest to determine the rate losses resulting from the use of
  correlated codes, defined as $\Delta_1 = I(X_1;Y_1)-I(X_1;Y_1|U_2)$,
  $\Delta_2 = I(X_2;Y_2)-I(X_2;Y_2|U_1)$, and
  $\Delta_0 = I(X_1;Y_1)+I(X_2;Y_2)-I(X_1X_2;Y_1Y_2)$.  Straightforward
  manipulations show that $\Delta_1 = I(Y_1;U_2)$, $\Delta_2 = I(Y_2;U_1)$,
  and $\Delta_0 = I(Y_1;Y_2)$.
\item Since $\Delta_i\geq 0$, $i\in\{0,1,2\}$ (mutual information is
  always nonnegative), we conclude that the region of achievable rates
  given by Corollary~\ref{cor:Machievable} is contained in the region
  defined by Corollary~\ref{cor:indep}.  Furthermore, we find that the
  rate loss terms have a simple, intuitive interpretation: $\Delta_0$
  is the loss in sum rate due to the dependencies between the outputs
  of different channels, and $\Delta_1$ (or $\Delta_2$) represent the
  rate loss due to the dependencies between the outputs of channel $1$
  (or $2$) and the source transmitted over channel $2$ (or $1$).  All
  these terms become zero if, instead of using correlated codes, we fix
  $p(x_1)p(x_2)$ and remove the correlation between the source blocks
  before transmission over the channels.
\end{itemize}
At first glance, this observation may seem somewhat surprising, since
the problem addressed by Corollary~\ref{cor:indep} is a special case
of the multiple access channel with correlated sources considered
in~\cite{CoverGS:80}, where it is shown that in the general case
correlated codes outperform the concatenation of Slepian-Wolf codes
(independent codewords) and optimal channel codes.  The crucial
difference between the two problems is the presence (or absence)
of interference in the channel.  Albeit somewhat informally, we can
state that correlated codes are advantageous when the transmitted
codewords are combined in the channel through interference, which
is obviously not the case in our problem.  Practical code constructions
built around this observation have been reported in~\cite{BarrosTL:03}.

\section{Conclusions}
\label{sec:conclusions}

\subsection{Summary}

In this paper we have considered the problem of encoding a set of
distributed correlated sources for delivery to a single data collector
node over a network of DMCs.  For this setup we were able to obtain
single-letter information theoretic conditions that provide an exact
characterization of the admissibility problem.  Two important conclusions
follow from the achievability proof:
\begin{itemize}
\item Separate source/channel coding is optimal in any network with one
  sink in which interference is dealt with at the MAC layer by creating
  independent links among nodes.
\item In such networks, the properties of Shannon information are
  exactly identical to those of water in pipes -- information is a
  flow.
\end{itemize}

\subsection{Discussion}

A few interesting observations follow from our results:

\begin{itemize}
\item It is a well known fact that turning a multiple access channel
  into an array of orthogonal channels by using a suitable MAC protocol
  is a suboptimal strategy in general, in the sense that the set of
  rates that are achievable with orthogonal access is strictly contained
  in the Ahlswede-Liao capacity region~\cite[Ch.\ 14.3]{CoverT:91}.
  However, despite its inherent suboptimality, there are strong economic
  incentives for the deployment of networks based on such technologies,
  related to the low complexity and cost of existing solutions, as well
  as experience in the fabrication and operation of such systems.  As
  a result, most existing standard implementations we are aware of
  (e.g., the IEEE 802.11 and 802.15.* families, or Bluetooth), are
  based on variants of protocols like TDMA/FDMA/CDMA or Aloha, that
  treat interference among users as noise or collisions, and deal with
  it by creating orthogonal links.  We feel therefore that some of the
  interest in our results stems from the fact that they provide a thorough
  analysis for what we deem to be, with high likelihood, the vast majority
  of wireless communication networks to be deployed for the foreseeable
  future.
\item A basic question follows from the results in this paper: when
  exactly does Shannon information act like a classical flow in a network
  setup?  In this paper, we showed that far more often than common wisdom
  would suggest: 
  for {\em any} network made up of independent links and one sink,
  Shannon information is a flow.  The assumption of independence among
  channels is crucial, since well known counterexamples hold without
  it~\cite{CoverGS:80}.  But, as argued before, far from being just some
  technical assumption needed for the theory to hold, independent channels
  arise naturally in practical applications.  In establishing the flow
  properties of information, we showed how some well understood network
  flow tools can be applied to address network design problems that
  have traditionally been difficult to deal with using standard tools
  in network information theory, and we illustrated this with a simple
  example involving optimal routing.  In particular we showed that, at
  least from an information theoretic point of view, there is little
  justification for the common practice of designing {\em trees} for
  collecting data picked up by a sensor network, thus opening up
  interesting problems of protocol design.
\item In retrospect, perhaps the results we prove in this paper should
  not have been surprising.  In the context of two-terminal networks, we
  do know the following:
  \begin{itemize}
  \item Feedback does not increase the capacity.  Therefore, the capacity
    of individual links is unaffected by the ability of our codes to
    establish a conference mechanism among nodes.
  \item Compression rates are not reduced by explicit cooperation, as it
    follows from the Slepian-Wolf theorem: the minimum rate required to
    communicate $U_1$ to a decoder that has access to side-information
    $U_0$ is $H(U_1|U_0)$, and knowledge of $U_0$ does not reduce the
    rates needed for coding $U_1$.  Therefore, the amount of information
    that needs to flow through our network is not reduced either by the
    ability of nodes to establish conferences.
  \end{itemize}
  Of course the statements above only hold for individual links, and a
  proof was needed to carry that intuition to the general network setup
  considered in this work.  But those observations we think are the
  key to understanding why our results hold.
\end{itemize}

\subsection{Future Work}

After having established coding theorems for the problem of network
information flow with correlated sources, a natural question that arises:
what if, in a given scenario, $\mathcal{R}=\emptyset$?  In that case,
the best we can hope for is to reconstruct an {\em approximation} to
the original source message --- and the answer is given by rate-distortion
theory~\cite{Berger:71}.  The rate-distortion formulation of our
problem in the case of non-cooperating encoders is equivalent to the
well known (and still open) {\em Multiterminal Source Coding}
problem~\cite{Berger:78}.  Our current efforts are focused on completing
work on the rate/distortion problem, and on fully developing the ideas
outlined in Section~\ref{sec:algorithmic-issues} (e.g., to deal with
problems of the type considered in~\cite{Chiang:05}).

\section*{Acknowledgements}

The authors most gratefully acknowledge discussions with Neri Merhav,
whose insightful comments on an earlier version of this manuscript led
to substantial improvements, as well as the valuable feedback from all
reviewers (and particularly from reviewer B).  They also wish to thank
Toby Berger and Te Sun Han for helpful discussions, and Joachim Hagenauer
for financial support without which they would have not been able to
work together.  The second author is also grateful to Mung Chiang, Eric
Friedman, \'Eva Tardos and Sergio Verd\'u, for useful discussions and
feedback on this work.

\appendix

\subsection{Converse Proof for Theorem~\ref{thm:main}}
\label{app:proof-converse-mcoop}

\subsubsection{Preliminaries}

Assume there exists a sequence of codes such that the decoder at $v_0$
is capable of producing a perfect reconstruction of blocks of $N$ snapshots
${\bf U} = [U_0^NU_1^N...U_M^N]$, with $P_e^{(N)}\to 0$ as $N\to\infty$.
Consider now decoding $L$ blocks of $N$ snapshots (indexed by $l=0...L-1$):
\begin{itemize}
\item The $1$-st block of snapshots ($l=0$) is computed based on
  messages $Y_{i0}^N$ received by $v_0$ from all nodes $v_i$ at
  times $kN$ ($k=0\,...\,W\!-\!1$).
\item The $2$-nd block of snapshots ($l=1$) is computed based on
  messages $Y_{i0}^N$ received by $v_0$ from all nodes $v_i$ at
  times $kN$ ($k=1\,...\,W$).
\item[] $\vdots$
\item The $L$-th block of snapshots ($l=L-1$) is computed based on
  messages $Y_{i0}^N$ received by $v_0$ from all nodes $v_i$ at
  times $kN$ ($k=L\!-\!1\,...\,W\!+\!(L\!-\!2)$).
\end{itemize}
Thus, we regard the network as a {\em pipeline}, in which ``packets''
(i.e., blocks of $N$ source symbols injected by each source) take
$NW$ units of time to flow, and each source gets to inject $L$ packets
total.  We are interested in the behavior of this pipeline in the
regime of large $L$.

For any fixed $L$, the probability of {\em at least one} of the $L$ blocks
being decoded in error is $P_e^{(LN)} = 1-(1-P_e^{(N)})^L$.  Thus, from the
existence of a code with low {\em block} probability of error we
can infer the existence of codes for which the probability of error
for the entire pipeline is low as well, by considering a large enough
block length $N$.

We begin with Fano's inequality. If there
is a suitable code as defined in the problem statement, then we must
have
\begin{equation}
  H(U_1^{LN}U_2^{LN}\dots U_M^{LN}
    | \hat{U}_1^{LN}\hat{U}_2^{LN}\dots \hat{U}_M^{LN})
  \;\; \leq \;\;
  P_e^{(LN)} \log\left(|{\mathcal U}_1^{LN}\!\times{\mathcal U}_2^{LN}
                       \!\times\dots\times{\mathcal U}_M^{LN}|\right)
                       + h(P_e^{(LN)}),
  \label{eq:fano2}
\end{equation}
where $h(\cdot)$ denotes the binary entropy function, and
$\hat U_{i}^{LN}=(\hat U_{i}^N(1),\hat U_{i}^N(2),\dots,\hat U_{i}^N(L))$
denotes $L$ blocks of $N$ snapshots reconstructed at $v_0$.
For convenience, we define also
\[ \delta(P_e^{(LN)}) \;\; = \;\;
   \left(P_e^{(LN)}\log\left(|{\mathcal U}_1^{LN}\times{\mathcal U}_2^{LN}
   \times\dots\times{\mathcal U}_M^{LN}|\right)+h(P_e^{(LN)})\right)/LN.
\]
It follows from eqn.~(\ref{eq:fano2}) that
\begin{eqnarray*}
\lefteqn{H(U_1^{LN}U_2^{LN}\dots U_M^{LN}|U_0^{LN}Y_{10}^{BN}Y_{20}^{BN}\dots Y_{M0}^{BN})} \\
 &\stackrel{(a)}{=} & H(U_1^{LN}U_2^{LN}\dots U_M^{LN}|U_0^{LN}Y_{10}^{BN}Y_{20}^{BN}\dots
          Y_{M0}^{BN}\hat{U}_1^{LN}\hat{U}_2^{LN}\dots \hat{U}_M^{LN})\\
  & \leq & H(U_1^{LN}U_2^{LN}\dots U_M^{LN}|\hat{U}_1^{LN}\hat{U}_2^{LN}\dots \hat{U}_M^{LN}) \\
  & \leq & LN \delta(P_e^{(LN)}),
\end{eqnarray*}
where $Y_{ij}^{BN}=(Y_{ij}^N(1),Y_{ij}^N(2),\dots,Y_{ij}^N(B))$ denotes
$B=W+(L-1)$ blocks of $N$ channel outputs observed by node $v_j$ while
communicating with node $v_i$, and (a) follows from the fact that the
estimates $\hat U_i^{LN}$, $i=1\dots M$, are functions of $U_0^{LN}$ and
of the received channel outputs $Y_{i0}^{BN}$, $i=1\dots M$.  From the
chain rule for entropy, from the fact that conditioning does not increase
entropy, and for any $S\subseteq {\cal M}=\{0...M\}$, $S\neq\emptyset$,
$0\in S^c$, it follows that
\begin{equation}
\label{eq:ineq}
 H(U^{LN}_S|U_{S^c}^{LN}Y_{S\to S^c}^{BN}Y_{S^c\to S^c}^{BN})
 \;\; \leq \;\;
 H(U^{LN}_S|U_{S^c}^{LN}Y_{S\to 0}^{BN}Y_{S^c\backslash\{0\}\to 0}^{BN})
 \;\; \leq \;\;
 LN \delta(P_e^{(LN)}).
\end{equation}
Let the set of $B$ codewords sent by
the nodes in a subset $A$ to the nodes in a subset $D$ be
\[X_{A\to D}^{BN}=\{X_{ij}^{BN}:i\in A \textrm{ and } j\in D\},\]
and, likewise, the corresponding channel outputs be denoted as
\[Y_{A\to D}^{BN}=\{Y_{ij}^{BN}:i\in A \textrm{ and } j\in D\}.\]

We will make use of the following lemmas.

\begin{lemma}\label{lemma:mi}
Let $X_{S\to S^c}$ be a set of channel inputs and $Y_{S\to S^c}$ be
a set of channel outputs of an array of independent channels
$\{{\cal X}_{ij},p_{ij}(y|x),{\cal Y}_{ij}\}$, $\forall i\in S$
and $\forall j\in S^c$.  Then,
\begin{equation}\label{eq:lemma}
I(X_{S\to S^c};Y_{S\to S^c})\leq \sum_{{i\in S},{j\in S^c}} I(X_{ij};Y_{ij}).
\end{equation}
\end{lemma}
\begin{proof}
Without loss of generality, assume that $S=\{1,\dots, x_0\}$ and
$S^c=\{x_0+1,\dots, M\}$.  From the definition of mutual information, it
follows that
\begin{eqnarray*}
I(X_{S\to S^c};Y_{S\to S^c})&=&H(Y_{S\to S^c})-H(Y_{S\to S^c}|X_{S\to S^c}).
\end{eqnarray*}
Expanding the first term on the right handside, we get
\begin{eqnarray*}
H(Y_{S\to S^c})&=&H(Y_{1\to S^c}Y_{2\to S^c}\dots Y_{x_0\to S^c})\\
&\leq& \sum_{i\in S} H(Y_{i\to S^c})\\
&=& \sum_{i\in S} H(Y_{i\to x_0+1}Y_{i\to x_0+2}\dots Y_{i\to M})\\
&\leq& \sum_{i\in S,j\in S^c} H(Y_{ij})\\
\end{eqnarray*}
Similarly, the second term reduces to
\begin{eqnarray*}
\lefteqn{H(Y_{S\to S^c}|X_{S\to S^c})}\\
&=&H(Y_{1\to S^c}Y_{2\to S^c}\dots Y_{x_0\to S^c}| X_{1\to S^c}X_{2\to S^c}\dots X_{x_0\to S^c})\\
&=& H(Y_{1\to S^c}| X_{1\to S^c}X_{2\to S^c}\dots X_{x_0\to S^c})+\sum_{i=2}^{x_0} H(Y_{i\to S^c}| X_{1\to S^c}X_{2\to S^c}\dots X_{x_0\to S^c}Y_{1\to S^c}\dots Y_{i-1\to S^c})\\
&=&  H(Y_{1\to S^c}| X_{1\to S^c})+\sum_{i=2}^{x_0} H(Y_{i\to S^c}| X_{i\to S^c})\\
&=& \sum_{i\in S}  H(Y_{i\to S^c}| X_{i\to S^c})\\
&=& \sum_{i\in S}  H(Y_{i\to x_0+1}Y_{i\to x_0+2}\dots Y_{i\to M}| X_{i\to x_0+1}X_{i\to x_0+2}\dots X_{i\to M})\\
&=& \sum_{i\in S} \bigg(H(Y_{i\to x_0+1}| X_{i\to x_0+1}X_{i\to x_0+2}\dots X_{i\to M})\\&& +\!\!\sum_{j=x_0+2}^M  H(Y_{i\to j}| X_{i\to x_0+1}X_{i\to x_0+2}\dots X_{i\to M})Y_{i\to x_0+1}\dots Y_{i\to j-1})\bigg)\\
&=& \sum_{i\in S} \bigg(H(Y_{i\to x_0+1}| X_{i\to x_0+1})+ \sum_{j=x_0+2}^M  H(Y_{i\to j}| X_{i\to j})\bigg)\\
&=& \sum_{i\in S,j\in S^c} H(Y_{ij}| X_{ij}).
\end{eqnarray*}
Combining the two expressions, we get
\[
  I(X_{S\to S^c};Y_{S\to S^c})
    \;\; \leq \;\; \sum_{i\in S,j\in S^c} H(Y_{ij})-H(Y_{ij}|X_{ij})
    \;\; = \;\; \sum_{i\in S,j\in S^c} I(X_{ij};Y_{ij}),
\]
thus proving the lemma.
\end{proof}

\begin{lemma}
\label{lemma:chain1}
$U_S^{LN}\rightarrow(U_{S^c}^{LN}Y_{S\to S^c}^{BN})\rightarrow
Y_{S^c\to S^c}^{BN}$ forms a Markov chain.
\end{lemma}
\begin{proof}
We begin by expanding $p(u_S^{LN}u_{S^c}^{LN}y_{S\to
S^c}^{BN}y_{S^c\to S^c}^{BN})$ according to
\begin{eqnarray*}
p(u_S^{LN}u_{S^c}^{LN}y_{S\to S^c}^{BN}y_{S^c\to S^c}^{BN})
&=&p(u_{S}^{LN}) \cdot p(u_{S^c}^{LN}y_{S\to S^c}^{BN}|u_S^{LN})
\cdot p(y_{S^c\to S^c}^{BN}|u_S^{LN}u_{S^c}^{LN}y_{S\to S^c}^{BN}).
\end{eqnarray*}
To prove that $U_S^{LN}$ can be removed from the last factor in
the previous expression, we will use an induction argument on the
length of the pipeline, $L$, and window sizes, $K$ and $T$.

Fix $(S,S^c)$ and $i,j\in S^c$. Let $L=K=T=1$. The encoding functions
produce $g_{ij}(U_i^N)=X_{i\to j}^N$, which result in the channel
outputs $Y_{i\to j}^N$ after transmission over the DMC between
nodes $i$ and $j$. In shorthand, we write
\[ g_{ij}(U_i^N)
   \;\; = \;\; X_{i\to j}^N
   \;\; \stackrel{\textrm{\tiny DMC}}{\longrightarrow} \;\; Y_{i\to j}^N.
\]
Thus, the first block of channel inputs
$X_{S^c\to S^c}^{1\dots N}$
generated in the node set $S_c$ depends only on source symbols
 $U_{S^c}^{1\dots N}$ available in $S_c$. Moreover,
since the channels are DMCs, the
channel outputs depend only on the channel inputs.
Thus, we conclude that $U_{S}^{1\dots N}$ and $Y_{S^c\to S^c}^{1\dots N}$
are independent given  $U_{S^c}^{1\dots N}$.

Since we consider a pipeline of length $L=1$, there are no more blocks
to inject, but not all data may have arrived to destination, so we have
to allow for a few ($W$, to be precise) extra transmissions.  By ``flushing
the pipeline'', we have
\[ g_{ij}(Y_{S\to i}^{1\dots N}Y_{S^c\to i}^{1\dots N})
   \;\; = \;\; X_{i\to j}^{N+1...2N}
   \;\; \stackrel{\textrm{\tiny DMC}}{\longrightarrow} \;\;
   Y_{i\to j}^{N+1\dots 2N}.
\]
It follows that $Y_{S^c\to S^c}^{N+1\dots 2N}$ is independent of
$U_S^{1\dots N}$ given $Y_{S\to S^c}^{1\dots N}$ and $U_{S^c}^{1\dots N}$.
Similarly, we have
\[ g_{ij}(Y_{S\to i}^{(W-2)N+1\dots(W-1)N}
          Y_{S^c\to i}^{(W-2)N+1\dots (W-1)N})
   \;\; = \;\; X_{i\to j}^{(W-1)N+1\dots WN}
   \;\; \stackrel{\textrm{\tiny DMC}}{\longrightarrow} \;\;
   Y_{i\to j}^{(W-1)N+1\dots WN},
\]
from which we conclude that
$Y_{S^c\to S^c}^{(W-1)N+1\dots WN}$ is independent of $U_S^{1\dots N}$
given $Y_{S\to S^c}^{(W-2)N+1\dots (W-1)N}$ and $U_{S^c}^{1\dots N}$.
Thus, for $K=T=L=1$, and $W$ arbitrary,\footnote{Since $W$ is the delay
used to allow data to flow to the destination, it would not be reasonable
to perform induction on $W$ for a given fixed network. Instead we take
$W$ as a parameter, which must be greater or equal to the diameter of
the network.} the Markov chain in the lemma holds (with $B=L+W-1$).

To proceed with the inductive proof, we still take $K=T=1$, $(S,S^c)$
fixed, $i,j\in S^c$, but $L$ is now arbitrary.  By inductive hypothesis,
we have the following Markov chain
\[ U_S^{(L-1)N}
   \;\; \rightarrow \;\; (U_{S^c}^{(L-1)N}Y_{S\to S^c}^{(B-1)N})
   \;\; \rightarrow \;\; Y_{S^c\to S^c}^{(B-1)N}.\]
Encoding and transmission of the last block of each source yields
\[ g_{ij}(U_i^{(L-1)N+1...LN}Y_{S\to i}^{(L-1)N+1\dots LN}
   Y_{S^c\to i}^{(L-1)N+1\dots LN})
   \;\; = \;\; X_{i\to j}^{LN+1\dots (L+1)N}
   \;\; \stackrel{\textrm{\tiny DMC}}{\longrightarrow} \;\;
   Y_{i\to j}^{LN+1\dots (L+1)N},
\]
such that for the last block, we have that
\[ U_S^{LN} \;\; \rightarrow \;\; (U_{S^c}^{LN}Y_{S\to S^c}^{(L+1)N})
   \;\; \rightarrow \;\; Y_{S^c\to S^c}^{(L+1)N}.\]
This is not yet the sought Markov chain, as we still need to flush the
pipe.  But similarly to how it was done for the base case of this inductive
argument, we have that
\[\begin{array}{ccccc}
  g_{ij}(Y_{S\to i}^{LN+1\dots (L+1)N}Y_{S^c\to i}^{LN+1\dots (L+1)N})
  & = & X_{i\to j}^{(L+1)N+1\dots (L+2)N}
  & \stackrel{\textrm{\tiny DMC}}{\longrightarrow} &
   Y_{i\to j}^{(L+1)N+1\dots (L+2)N}, \\
  & & \vdots & &\\
  g_{ij}(Y_{S\to i}^{(B-2)N+1\dots (B-1)N}Y_{S^c\to i}^{(B-2)N+1\dots (B-1)N})
  & = & X_{i\to j}^{(B-1)N+1\dots BN}
  & \stackrel{\textrm{\tiny DMC}}{\longrightarrow} &
   Y_{i\to j}^{(B-1)N+1\dots BN},
\end{array}\]
and therefore, now yes, we have that
$Y_{S^c\to S^c}^{BN}$ is independent of $U_S^{1\dots N}$
given $Y_{S\to S^c}^{BN}$ and $U_{S^c}^{1\dots N}$.

The proof of the lemma is completed by performing the exact same
induction steps on $K$ and $T$ as done on $L$.  For brevity, those
same steps are omitted from this proof.
\end{proof}

\subsubsection{Main Proof}

We now take an arbitrary non-empty subset $S \subseteq {\cal M}=\{0...M\}$,
$S\neq\emptyset$, $0\in S^c$. and start by bounding $H(U_S^{LN})$ according
to
\begin{eqnarray*}
H(U_S^{LN})
  &=& I\big(U_S^{LN};U_{S^c}^{LN}Y_{S\to S^c}^{BN}Y_{S^c\to S^c}^{BN}\big)\;+\;
      H\big(U_S^{LN}|U_{S^c}^{LN}Y_{S\to S^c}^{BN}Y_{S^c\to S^c}^{BN}\big) \\
  &\stackrel{(a)}{\leq}&
    I\big(U_S^{LN};U_{S^c}^{LN}Y_{S\to S^c}^{BN}Y_{S^c\to S^c}^{BN}\big)
    \;+\; LN\delta(P_e^{(LN)}) \\
  &=& I\big(U_S^{LN};U_{S^c}^{LN}\big)
    \;+\;I(U_S^{LN};Y_{S\to S^c}^{BN}|U_{S^c}^{LN})
    \;+\;I(U_S^{LN};Y_{S^c\to S^c}^{BN}|U_{S^c}^{LN}Y_{S\to S^c}^{BN})
    \:+\;LN\delta(P_e^{(LN)}),
\end{eqnarray*}
where (a) follows from~(\ref{eq:ineq}).  From Lemma~\ref{lemma:chain1}, we
have that $I(U_S^{LN};Y_{S^c\to S^c}^{BN}|U_{S^c}^{LN}Y_{S\to S^c}^{BN}) = 0$,
and so we get
\begin{equation}
H(U_S^{LN}) \;\; \leq \;\;
  I(U_S^{LN};U_{S^c}^{LN}) \; + \; I(U_S^{LN};Y_{S\to S^c}^{BN}|U_{S^c}^{LN})
  \; + \; LN\delta(P_e^{(LN)}).
\label{eq:almostend}
\end{equation}

Developing the second term on the right handside yields:
\begin{eqnarray*}
\lefteqn{I(U_S^{LN};Y_{S\to S^c}^{BN}|U_{S^c}^{LN})} \\
& = & \sum_{k=1}^{BN}I(U_S^{LN};Y_{S\to S^c}(k)|U_{S^c}^{LN}Y_{S\to S^c}^{k-1})\\
& \leq & \sum_{k=1}^{BN}I(U_S^{LN};Y_{S\to
S^c}(k)|U_{S^c}^{LN}Y_{S\to
S^c}^{k-1}) +\sum_{k=1}^{BN}I(X_{S\to
S^c}(k);Y_{S\to
S^c}(k)|U_{S^c}^{LN}Y_{S\to
S^c}^{k-1}U_S^{LN})
\\
&=&\sum_{k=1}^{BN}I(X_{S\to S^c}(k)U_S^{LN};Y_{S\to
S^c}(k)|U_{S^c}^{LN}Y_{S\to
S^c}^{k-1})
\\
&=&\sum_{k=1}^{BN}I(X_{S\to S^c}(k);Y_{S\to S^c}(k)|U_{S^c}^{LN}Y_{S\to S^c}^{k-1}) +I(U_S^{LN};Y_{S\to S^c}(k)|U_{S^c}^{LN}Y_{S\to S^c}^{k-1}X_{S\to S^c}(k))\\
\end{eqnarray*}\begin{eqnarray*}
&\stackrel{(a)}{=}&\sum_{k=1}^{BN}I(X_{S\to S^c}(k);Y_{S\to S^c}(k)|U_{S^c}^{LN}Y_{S\to S^c}^{k-1})\\
&=&\sum_{k=1}^{BN}H(Y_{S\to
S^c}(k)|U_{S^c}^{LN}Y_{S\to
S^c}^{k-1}) -
H(Y_{S\to S^c}(k)|U_{S^c}^{LN}Y_{S\to S^c}^{k-1}X_{S\to S^c}(k))\\
&\stackrel{(b)}{=}&\sum_{k=1}^{BN}H(Y_{S\to S^c}(k)|U_{S^c}^{LN}Y_{S\to S^c}^{k-1})-H(Y_{S\to S^c}(k)|X_{S\to S^c}(k))\\
&\stackrel{(c)}{\leq}&\sum_{k=1}^{BN}H(Y_{S\to
S^c}(k))-H(Y_{S\to
S^c}(k)|X_{S\to S^c}(k))\\
&=&\sum_{k=1}^{BN}I(X_{S\to S^c}(k);Y_{S\to S^c}(k))\\
&\stackrel{(d)}{\leq}&\sum_{k=1}^{BN}\sum_{i\in S, j\in S^c} I(X_{ij}(k);Y_{ij}(k))\\
&=&\sum_{i\in S, j\in S^c} \sum_{k=1}^{BN} I(X_{ij}(k);Y_{ij}(k))\\
&\leq&\sum_{i\in S, j\in S^c}  BNC_{ij}
\end{eqnarray*}
where we use the following arguments:
\begin{itemize}
\item[(a)] given the channel inputs  $X_{S\to S^c}(i)$ the
  channel outputs $Y_{S\to S^c}(i)$ are independent of all
  other random variables;
\item[(b)] same as (a);
\item[(c)] conditioning does not increase the entropy;
\item[(d)] direct application of lemma \ref{lemma:mi}.
\end{itemize}
Substituting in (\ref{eq:almostend}) yields
\begin{eqnarray*}
 H(U_S^{LN}) &\leq& I(U_S^{LN};U_{S^c}^{LN}) + \sum_{i\in S, j\in S^c} BNC_{ij}
                    + LN\delta(P_e^{(LN)}).
\end{eqnarray*}
Using the fact that the sources are drawn i.i.d., this last expression
can be rewritten as
\[
  LNH(U_S) \;\;\leq\;\; LNI(U_S;U_{S^c})
               + \sum_{i\in S, j\in S^c} BNC_{ij}
                + LN\delta(P_e^{(LN)}),
\]
or equivalently,
\[
H(U_S|U_{S^c}) \;\;\leq\;\;  \frac{B}{L}\sum_{i\in S, j\in S^c}
C_{ij} + \delta(P_e^{(LN)})
\;\;\leq\;\; \frac{(W+L-1)}{L} \sum_{i\in S, j\in S^c} C_{ij}+\delta(P_e^{(LN)})\\
\]
Finally, we observe that this inequality holds for all finite values
of $L$.  Thus, it must also be the case that
\begin{eqnarray*}
H(U_S|U_{S^c})
  & < & \inf_{L=1,2,...}
           \frac{(W+L-1)}{L} \sum_{i\in S, j\in S^c} C_{ij}+\delta(P_e^{(LN)}) \\
  & = &    \sum_{i\in S, j\in S^c} C_{ij}+\delta(P_e^{(LN)}).
\end{eqnarray*}
But since $\delta(P_e^{(LN)})$ goes to zero as $P_e^{(N)}\rightarrow 0$,
we get
\[ H(U_S|U_{S^c}) \;\;<\;\; \sum_{i\in S, j\in S^c} C_{ij}, \]
thus concluding the proof.  \tend


\end{document}